\documentclass[preprint2]{aastex61}%This is for two column
\pdfoutput=1

\usepackage{amsmath,multirow}
\newcommand\aastex{AAS\TeX}

\shorttitle{\aastex\ MWL observations of the blazar BL~Lacertae: a new fast TeV gamma-ray flare }
\shortauthors{VERITAS et al.}

\begin{document}

\title{Multiwavelength observations of the blazar BL~Lacertae: a new fast TeV gamma-ray flare}

\correspondingauthor{Qi Feng, Olivier Hervet, Svetlana Jorstad}
%\email{qi.feng2@mcgill.ca}
\email{qifeng@nevis.columbia.edu, ohervet@ucsc.edu, jorstad@bu.edu}

%\correspondingauthor{Olivier Hervet}
%\email{ohervet@ucsc.edu}

\author{A.~U.~Abeysekara}
\affiliation{Department of Physics and Astronomy, University of Utah, Salt Lake City, UT 84112, USA}

\author{W.~Benbow}
\affiliation{Fred Lawrence Whipple Observatory, Harvard-Smithsonian Center for Astrophysics, Amado, AZ 85645, USA}

\author{R.~Bird}
\affiliation{Department of Physics and Astronomy, University of California, Los Angeles, CA 90095, USA}

\author{T.~Brantseg}
\affiliation{Department of Physics and Astronomy, Iowa State University, Ames, IA 50011, USA}

\author{R.~Brose}
\affiliation{Institute of Physics and Astronomy, University of Potsdam, 14476 Potsdam-Golm, Germany}
\affiliation{DESY, Platanenallee 6, 15738 Zeuthen, Germany}

\author{M.~Buchovecky}
\affiliation{Department of Physics and Astronomy, University of California, Los Angeles, CA 90095, USA}

\author{J.~H.~Buckley}
\affiliation{Department of Physics, Washington University, St. Louis, MO 63130, USA}

\author{V.~Bugaev}
\affiliation{Department of Physics, Washington University, St. Louis, MO 63130, USA}

\author{M.~P.~Connolly}
\affiliation{School of Physics, National University of Ireland Galway, University Road, Galway, Ireland}

\author{W.~Cui}
\affiliation{Department of Physics and Astronomy, Purdue University, West Lafayette, IN 47907, USA}
\affiliation{Department of Physics and Center for Astrophysics, Tsinghua University, Beijing 100084, China.}

\author{M.~K.~Daniel}
\affiliation{Fred Lawrence Whipple Observatory, Harvard-Smithsonian Center for Astrophysics, Amado, AZ 85645, USA}

\author{A.~Falcone}
\affiliation{Department of Astronomy and Astrophysics, 525 Davey Lab, Pennsylvania State University, University Park, PA 16802, USA}

%\author[0000-0001-6674-4238]{Q.~Feng}
\author{Q.~Feng}
\affiliation{Physics Department, McGill University, Montreal, QC H3A 2T8, Canada}
\affiliation{Physics Department, Columbia University, New York, NY 10027, USA}
%\collaboration{(The VERITAS Collaboration)}

\author{J.~P.~Finley}
\affiliation{Department of Physics and Astronomy, Purdue University, West Lafayette, IN 47907, USA}

\author{L.~Fortson}
\affiliation{School of Physics and Astronomy, University of Minnesota, Minneapolis, MN 55455, USA}

\author{A.~Furniss}
\affiliation{Department of Physics, California State University - East Bay, Hayward, CA 94542, USA}

\author{G.~H.~Gillanders}
\affiliation{School of Physics, National University of Ireland Galway, University Road, Galway, Ireland}

\author{I.~Gunawardhana}
\affiliation{Department of Physics and Astronomy, University of Utah, Salt Lake City, UT 84112, USA}

\author{M.~H\"utten}
\affiliation{DESY, Platanenallee 6, 15738 Zeuthen, Germany}

\author{D.~Hanna}
\affiliation{Physics Department, McGill University, Montreal, QC H3A 2T8, Canada}

\author{O.~Hervet}
\affiliation{Santa Cruz Institute for Particle Physics and Department of Physics, University of California, Santa Cruz, CA 95064, USA}

\author{J.~Holder}
\affiliation{Department of Physics and Astronomy and the Bartol Research Institute, University of Delaware, Newark, DE 19716, USA}

\author{G.~Hughes}
\affiliation{Fred Lawrence Whipple Observatory, Harvard-Smithsonian Center for Astrophysics, Amado, AZ 85645, USA}

\author{T.~B.~Humensky}
\affiliation{Physics Department, Columbia University, New York, NY 10027, USA}

\author{C.~A.~Johnson}
\affiliation{Santa Cruz Institute for Particle Physics and Department of Physics, University of California, Santa Cruz, CA 95064, USA}

\author{P.~Kaaret}
\affiliation{Department of Physics and Astronomy, University of Iowa, Van Allen Hall, Iowa City, IA 52242, USA}

\author{P.~Kar}
\affiliation{Department of Physics and Astronomy, University of Utah, Salt Lake City, UT 84112, USA}

\author{M.~Kertzman}
\affiliation{Department of Physics and Astronomy, DePauw University, Greencastle, IN 46135-0037, USA}

\author{F.~Krennrich}
\affiliation{Department of Physics and Astronomy, Iowa State University, Ames, IA 50011, USA}

\author{M.~J.~Lang}
\affiliation{School of Physics, National University of Ireland Galway, University Road, Galway, Ireland}

\author{T.~T.Y.~Lin}
\affiliation{Physics Department, McGill University, Montreal, QC H3A 2T8, Canada}

\author{S.~McArthur}
\affiliation{Department of Physics and Astronomy, Purdue University, West Lafayette, IN 47907, USA}

\author{P.~Moriarty}
\affiliation{School of Physics, National University of Ireland Galway, University Road, Galway, Ireland}

\author{R.~Mukherjee}
\affiliation{Department of Physics and Astronomy, Barnard College, Columbia University, NY 10027, USA}

\author{S.~O'Brien}
\affiliation{School of Physics, University College Dublin, Belfield, Dublin 4, Ireland}

\author{R.~A.~Ong}
\affiliation{Department of Physics and Astronomy, University of California, Los Angeles, CA 90095, USA}

\author{A.~N.~Otte}
\affiliation{School of Physics and Center for Relativistic Astrophysics, Georgia Institute of Technology, 837 State Street NW, Atlanta, GA 30332-0430}

\author{N.~Park}
\affiliation{Enrico Fermi Institute, University of Chicago, Chicago, IL 60637, USA}

\author{A.~Petrashyk}
\affiliation{Physics Department, Columbia University, New York, NY 10027, USA}

\author{M.~Pohl}
\affiliation{Institute of Physics and Astronomy, University of Potsdam, 14476 Potsdam-Golm, Germany}
\affiliation{DESY, Platanenallee 6, 15738 Zeuthen, Germany}

\author{E.~Pueschel}
\affiliation{DESY, Platanenallee 6, 15738 Zeuthen, Germany}

\author{J.~Quinn}
\affiliation{School of Physics, University College Dublin, Belfield, Dublin 4, Ireland}

\author{K.~Ragan}
\affiliation{Physics Department, McGill University, Montreal, QC H3A 2T8, Canada}

\author{P.~T.~Reynolds}
\affiliation{Department of Physical Sciences, Cork Institute of Technology, Bishopstown, Cork, Ireland}

\author{G.~T.~Richards}
\affiliation{School of Physics and Center for Relativistic Astrophysics, Georgia Institute of Technology, 837 State Street NW, Atlanta, GA 30332-0430}

\author{E.~Roache}
\affiliation{Fred Lawrence Whipple Observatory, Harvard-Smithsonian Center for Astrophysics, Amado, AZ 85645, USA}

\author{C.~Rulten}
\affiliation{School of Physics and Astronomy, University of Minnesota, Minneapolis, MN 55455, USA}

\author{I.~Sadeh}
\affiliation{DESY, Platanenallee 6, 15738 Zeuthen, Germany}

\author{M.~Santander}
\affiliation{Department of Physics and Astronomy, Barnard College, Columbia University, NY 10027, USA}
\affiliation{Department of Physics and Astronomy, University of Alabama, Tuscaloosa, AL 35487, USA}

\author{G.~H.~Sembroski}
\affiliation{Department of Physics and Astronomy, Purdue University, West Lafayette, IN 47907, USA}

\author{K.~Shahinyan}
\affiliation{School of Physics and Astronomy, University of Minnesota, Minneapolis, MN 55455, USA}

\author{S.~P.~Wakely}
\affiliation{Enrico Fermi Institute, University of Chicago, Chicago, IL 60637, USA}

\author{A.~Weinstein}
\affiliation{Department of Physics and Astronomy, Iowa State University, Ames, IA 50011, USA}

\author{R.~M.~Wells}
\affiliation{Department of Physics and Astronomy, Iowa State University, Ames, IA 50011, USA}

\author{P.~Wilcox}
\affiliation{Department of Physics and Astronomy, University of Iowa, Van Allen Hall, Iowa City, IA 52242, USA}

\author{D.~A.~Williams}
\affiliation{Santa Cruz Institute for Particle Physics and Department of Physics, University of California, Santa Cruz, CA 95064, USA}

\author{B.~Zitzer}
\affiliation{Physics Department, McGill University, Montreal, QC H3A 2T8, Canada}

\collaboration{(The VERITAS Collaboration)}

\author{S.~G.~Jorstad} 
\affiliation{Institute for Astrophysical Research, Boston University, 725 Commonwealth Avenue, Boston, MA 02215, USA }
\affiliation{Astronomical Institute, St.Petersburg State University, Universitetskij Pr. 28, Petrodvorets, 198504 St.Petersburg, Russia}

\author{A.~P.~Marscher} 
\affiliation{Institute for Astrophysical Research, Boston University, 725 Commonwealth Avenue, Boston, MA 02215, USA }

\author{M.~L.~Lister} 
\affiliation{Purdue University, 525 Northwestern Avenue, West Lafayette, IN 47907, USA }

\author{Y.~Y.~Kovalev} 
\affiliation{Astro Space Center of Lebedev Physical Institute, Profsoyuznaya 84/32, 117997 Moscow, Russia }
\affiliation{Moscow Institute of Physics and Technology, Dolgoprudny, Institutsky per., 9, Moscow region, 141700, Russia}
\affiliation{Max-Planck-Institut f\"ur Radioastronomie, Auf dem H\"ugel 69, D-53121 Bonn, Germany }

\author{A.~B.~Pushkarev} 
\affiliation{Crimean Astrophysical Observatory, 98409 Nauchny, Crimea, Russia }
\affiliation{Astro Space Center of Lebedev Physical Institute, Profsoyuznaya 84/32, 117997 Moscow, Russia }

\author{T.~Savolainen} 
\affiliation{Aalto University Mets\"ahovi Radio Observatory, Mets\"ahovintie 114, FI-02540 Kylm\"al\"a, Finland}
\affiliation{Aalto University Department of Electronics and Nanoengineering, PL 15500, FI-00076 Aalto, Finland }
\affiliation{Max-Planck-Institut f\"ur Radioastronomie, Auf dem H\"ugel 69, D-53121 Bonn, Germany }

\author{I.~Agudo} 
\affiliation{Instituto de Astrof\'{\i}sica de Andaluc\'{\i}a (CSIC), Apartado 3004, E--18080 Granada, Spain }

\author{S.~N.~Molina}  
\affiliation{Instituto de Astrof\'{\i}sica de Andaluc\'{\i}a (CSIC), Apartado 3004, E--18080 Granada, Spain }

\author{J.~L.~G\'{o}mez}  
\affiliation{Instituto de Astrof\'{\i}sica de Andaluc\'{\i}a (CSIC), Apartado 3004, E--18080 Granada, Spain }

\author{V.~M.~Larionov}  
\affiliation{Astronomical Institute, St.Petersburg State University, Universitetskij Pr. 28, Petrodvorets, 198504 St.Petersburg, Russia}

\author{G.~A.~Borman}  
\affiliation{Crimean Astrophysical Observatory, 98409 Nauchny, Crimea, Russia}

\author{A.~A.~Mokrushina}
\affiliation{Astronomical Institute, St.Petersburg State University, Universitetskij Pr. 28, Petrodvorets, 198504 St.Petersburg, Russia}

\author{M. Tornikoski}
%  Affiliation 1)
\affiliation{Aalto University Mets\"ahovi Radio Observatory, Mets\"ahovintie 114, FI-02540 Kylm\"al\"a, Finland}

\author{A. L\"ahteenm\"aki}
%  Affiliations 1), 2), and 3)
%  Affiliation 1)
\affiliation{Aalto University Mets\"ahovi Radio Observatory, Mets\"ahovintie 114, FI-02540 Kylm\"al\"a, Finland}
%2)
%\affiliation{Aalto University Department of Electronics and Nanoengineering, P.O. BOX 15500, FI-00076 AALTO, Finland}
\affiliation{Aalto University Department of Electronics and Nanoengineering, PL 15500, FI-00076 Aalto, Finland }

%3)
\affiliation{Tartu Observatory Observatooriumi 161602 T\~{o}ravere, Estonia}

\author{W. Chamani}
%  Affiliations 1), 2)
%  Affiliation 1)
\affiliation{Aalto University Mets\"ahovi Radio Observatory, Mets\"ahovintie 114, FI-02540 Kylm\"al\"a, Finland}
%2)
%\affiliation{Aalto University Department of Electronics and Nanoengineering, P.O. BOX 15500, FI-00076 AALTO, Finland}
\affiliation{Aalto University Department of Electronics and Nanoengineering, PL 15500, FI-00076 Aalto, Finland }

\author{S. Enestam}
%  Affiliations 1), 2)
%  Affiliation 1)
\affiliation{Aalto University Mets\"ahovi Radio Observatory, Mets\"ahovintie 114, FI-02540 Kylm\"al\"a, Finland}
%2)
%\affiliation{Aalto University Department of Electronics and Nanoengineering, P.O. BOX 15500, FI-00076 AALTO, Finland}
\affiliation{Aalto University Department of Electronics and Nanoengineering, PL 15500, FI-00076 Aalto, Finland }

\author{S. Kiehlmann}
\affiliation{Owens Valley Radio Observatory, California Institute of Technology, Pasadena, CA 91125, USA}

\author{T. Hovatta}
\affiliation{Tuorla Observatory, Department of Physics and Astronomy, University of Turku, V\"ais\"al\"antie 20, 21500 Kaarina, Finland}

\author{P.~S.~Smith}
\affiliation{Steward Observatory, University of Arizona, Tucson, AZ 85716, USA}

\author{P. Pontrelli}
\affiliation{Santa Cruz Institute for Particle Physics and Department of Physics, University of California, Santa Cruz, CA 95064, USA}

\begin{abstract}
Combined with very-long-baseline interferometry measurements, the observations of fast TeV gamma-ray flares probe the structure and emission mechanism of blazar jets. 
However, only a handful of such flares have been detected to date, and only within the last few years have these flares been observed from lower-frequency-peaked BL~Lac objects and flat-spectrum radio quasars. 
We report on a fast TeV gamma-ray flare from the blazar BL~Lacertae observed by VERITAS, with a rise time of $\sim$2.3~hr and a decay time of $\sim$36~min. 
The peak flux above 200 GeV is $(4.2 \pm 0.6) \times 10^{-6} \;\text{photon} \;\text{m}^{-2}\; \text{s}^{-1}$ measured with a 4-minute-binned light curve, corresponding to $\sim$180\% of the flux which is observed from the Crab Nebula above the same energy threshold. 
Variability contemporaneous with the TeV gamma-ray flare was observed in GeV gamma-ray, X-ray, and optical flux, as well as in optical and radio polarization. 
Additionally, a possible moving emission feature with superluminal apparent velocity was identified in VLBA observations at 43 GHz, potentially passing the radio core of the jet around the time of the gamma-ray flare. 
We discuss the constraints on the size, Lorentz factor, and location of the emitting region of the flare, and the interpretations with several theoretical models which invoke relativistic plasma passing stationary shocks. 
\end{abstract}

%%%%%%%%%
% Intro
%%%%%%%%%
\section{Introduction} \label{sec:intro}

BL~Lac objects belong to a subclass of radio-loud active galactic nuclei (AGNs), known as blazars. They are characterized by featureless optical spectra, non-thermal broadband spectra, and rapid variability, which jointly suggest that their emission originates in relativistic jets closely aligned to our line of sight \citep[e.g.,][and references therein]{BlandfordRees78}. 

Fast variability at very high energies (100~GeV $\lesssim E_{\gamma} \lesssim$ 100~TeV; VHE), with timescales as short as a few minutes, has been observed in several blazars \citep[e.g.,][]{Gaidos96, Aharonian07, Albert07, Aleksic11}, including the prototypical BL~Lacertae \citep[VER~J2202+422;][]{Arlen13} located at redshift $z=0.069$ \citep{Miller77}. %EG2015IC310 
Long-term monitoring of BL~Lacertae %by multiple TeV gamma-ray instruments 
has led to no detection of the source in the TeV gamma-ray band by the current generation of instruments except during flaring episodes, when its flux has been observed to reach $>$100\% of the Crab Nebula flux (C.~U.) above 1 TeV in 1998 \citep{Neshpor01}, $\sim0.03$~C.~U. above 200~GeV in 2005 \citep{Albert07BLLac}, and most recently $\sim1.25$~C.~U. above 200~GeV with a short variability timescale of $13\pm4$ min in 2011 \citep{Arlen13}. 
%BL~Lacertae (also known as 1ES 2200+420) is an AGN located at a redshift of $z=0.069$ \citep{Miller78}. In 1998, the Crimean Observatory reported a detection of the source at $>$100\% of the Crab Nebula flux above 1 TeV \citep{Neshpor01}. Subsequently, the MAGIC collaboration reported another detection during an active state in 2005, but at a much lower flux level (only about 3\% of the Crab Nebula flux) \citep{Albert07BLLac}. Triggered by the elevated activities recently seen by the {\it Fermi} LAT \citep{Atel3368} and AGILE \citep{Atel3387} at GeV gamma-ray energies, as well as in the optical \citep{Atel3371}, near infra-red \citep{Atel3375}, and radio \citep{Atel3380} in 2011 May, we began to monitor BL~Lacertae more regularly at TeV gamma-ray energies with VERITAS. In this work, we report the detection of a rapid, intense VHE gamma-ray flare in the general direction of the source on MJD 55740 (2011 June 28), as well as the results from the MWL observations that were conducted around the time of this flare. 

The rapid gamma-ray variability observed in TeV blazars implies very compact emitting regions, as well as low gamma-ray attenuation by pair production on infrared/optical photons near the emission zone. 
%In a one-zone synchrotron self-Compton model, one of the most popular and the simplest blazar models \citep[e.g.,][]{Ghisellini98, Bottcher02}, the intrinsic pair-production opacity of a relativistic emission zone depends on its size and Doppler factor, and on the density of lower-energy photons. 
While a one-zone synchrotron self-Compton (SSC) model, one of the simplest blazar models \citep[e.g.,][]{Ghisellini98, Bottcher02}, has 
been effective at explaining emission from high-frequency-peaked BL Lac (HBL) objects, the intrinsic pair-production opacity of a relativistic emission zone in such a model depends on its size and Doppler factor, and on the density of lower-energy photons. 
Therefore, if the synchrotron photons are the main source of the lower-energy radiation, the emitting region must have a small size and/or a large Doppler factor so that the gamma rays can escape pair production. 
Alternatively, if an external photon field (e.g., the broad-line region; BLR) dominates the lower-energy radiation, it can cause substantial gamma-ray absorption. As a result, the emitting region is generally expected to be far away from the central region of the AGN, especially for flat-spectrum radio quasars (FSRQs) 
whose broad-line emission is relatively strong. 

BL~Lacertae was first classified as a low-frequency-peaked BL Lac (LBL) object as the synchrotron peak frequency of its spectral energy distribution (SED) was measured to be $2.2 \times 10^{14}$ Hz \citep{Sambruna99}, but was later reclassified as an intermediate-frequency-peaked BL Lac (IBL) object \citep{Ackermann11}. 
It has been reported that the SEDs of several IBLs/LBLs cannot be well described by a one-zone SSC model \citep[see][and references therein]{Hervet15}, and more complex models such as multi-zone SSC models or external-radiation Compton (ERC) models are needed. 
%

% or be sufficiently far away from the central broad-line region (BLR), whose radiation field can lead to significant attenuation of the gamma-ray flux. %
%We note that although the gamma-ray opacity induced by an external photon field does not depend on the Doppler factor, a high Doppler factor is necessary for an emission zone inside the BLR in the case of a gamma-ray luminosity dominated by the external inverse Compton process.

The large Doppler factor and/or distant downstream emitting region required by the observed fast TeV variability of blazars, together with the knotty jet structures (both moving and stationary) identified with high-resolution radio observations \citep[e.g.,][]{Cohen14}, can be explained consistently by theoretical models with multiple emitting zones that are either spatially or temporally separated, %or multiple particle populations, 
e.g., structured jets \citep{Ghisellini05}, jet deceleration \citep{Stern08}, jets in a jet \citep{Giannios09}, and plasma passing a standing shock \citep[e.g.,][]{Marscher14, Zhang14, Hervet16, Pollack16}. 

However, the details regarding the location and the mechanism of blazar emission are still not well understood \citep[e.g.,][]{Madejski16}. 
Simultaneous multiwavelength (MWL) observations can provide insights into the flaring mechanisms (e.g., leptonic or hadronic processes) 
of these objects, particularly at the wavelengths where SEDs often peak. 
%Particularly, the X-ray and gamma-ray measurements are of interest, as the spectral energy distributions (SEDs) of TeV blazars usually show two characteristic peaks in these bands \citep[e.g.][]{Fossati98}. However, it is unlikely that current X-ray instruments can respond fast enough to a sub-hour TeV gamma-ray flare. The chances of such a flare happening during pre-scheduled simultaneous X-ray and TeV gamma-ray observations are small, despite dedicated strategies were attempted \citep[e.g.][]{Abeysekara17}. 
%Such observations are challenging practically for fast flares at sub-hour timescales, 
In practice, such observations are limited in the case of fast flares at sub-hour timescales,
even with dedicated strategies \citep[e.g.,][]{Abeysekara17}. 
%Although strictly simultaneous data during rapid TeV gamma-ray flares of blazars are lacking, 
Nevertheless, contemporaneous radio data are often relevant because the radio variability timescale is usually much longer \citep[e.g.,][]{Rani13}. 
In particular, the evolution of polarization (both radio and optical) before and after a gamma-ray flare provides information about magnetic field structures of the jet, and therefore the activity of possible gamma-ray emitting regions \citep[e.g.,][]{Zhang14}. 

BL~Lacertae exhibits both stationary radio cores/knots and superluminal radio knots \citep{Lister13, Gomez16}. 
Possible associations between the variability of superluminal radio knots and gamma-ray flares have been investigated for BL~Lacertae \citep[e.g.,][]{Marscher08, Arlen13} and other blazars \citep[e.g.,][]{Rani14, Max-Moerbeck14}. 

On 2016 Oct 5, we observed BL~Lacertae at an elevated flux level with sub-hour variability with the Very Energetic Radiation Imaging Telescope Array System (VERITAS; %as part of the long-term-plan snapshot program 
see Section~\ref{sec:VERITAS}). 
A series of observations with the Very Long Baseline Array (VLBA) at 43~GHz and 15.4~GHz was performed %(and 86~GHz?) 
over a few months before and after the gamma-ray flare, revealing a possible knot structure emerging around the time of the TeV flare (see Section~\ref{sec:VLBA}).
In this work, we report on the results of the VERITAS, VLBA, and other MWL observations and discuss their implications. 
The cosmological parameters assumed throughout this paper are $\Omega_m=0.27$, $\Omega_{\Lambda}=0.73$, and $H_0=70\;\text{km}\;\text{s}^{-1}\;\text{Mpc}^{-1}$ \citep{Larson11}. At the redshift of BL~Lacertae, the luminosity distance and the angular size distance are $311$~Mpc and $273$~Mpc, respectively, and the angular scale is $1.3$~pc/mas. 

%%%%%%%%%
% Observations and data analysis
%%%%%%%%%
\section{Observations, Data Analysis, and Results}
\label{sec:obs_data}
%
%%%%%%%%%
\subsection{VERITAS}
\label{sec:VERITAS}
%\subsection{VHE gamma-ray observations with VERITAS}
%%%%%%%%%
%
VERITAS is an array of four imaging atmospheric-Cherenkov telescopes located in southern Arizona \citep[30$^\circ$ 40' N, 110$^\circ$ 57' W, 1.3 km above sea level;][]{Holder11}. It is sensitive to gamma rays in the energy range from 85 GeV to $>$30 TeV with an energy resolution of $\sim$15\% (at 1~TeV) and is capable of making a detection with a statistical significance of 5 standard deviations ($5\sigma$) of a point source of 0.01~C.~U. in $\sim$25~hr. 
%
%Each of the four telescopes is equipped with a 12-m diameter Davies-Cotton reflector comprising 355 identical mirror facets, 
%and a 499-pixel photomultiplier tube (PMT) camera covering a field of view of 3.5$^\circ$ at an angular resolution (68\% containment) of $\sim$0.1$^{\circ}$ (at 1~TeV). Coincident Cherenkov signals from at least two out of the four telescopes are required to trigger an array-wide read-out of the PMT signals. 

BL~Lacertae was observed at an elevated TeV gamma-ray flux by VERITAS on 2016 Oct 5 as part of an ongoing monitoring program, %the long-term-plan snapshot program \citep{Benbow16}, 
and follow-up observations were immediately instigated %instantaneously followed up %by VERITAS 
based on a real-time analysis. The total exposure of these observations amounts to 153.5 min after data-quality selection, with zenith angles ranging between 11$^\circ$ and 30$^\circ$. 
%The data were analyzed using two independent analysis packages \citep{Cogan08, Daniel08}. 
%A pre-determined set of cuts optimized for lower-energy showers were used in the analysis \citep[see e.g.][]{Archambault14}, which yield a detection with a statistical significance of $70.7\sigma$ for all the data of the night. The time-averaged integral flux above 200~GeV is $(2.24 \pm 0.06) \times 10^{-6} \;\text{photons} \;\text{m}^{-2}\; \text{s}^{-1}$.
The data were analyzed using two independent analysis packages \citep{Cogan08, Daniel08} 
and a pre-determined set of cuts optimized for lower-energy showers \cite[see e.g.,][]{Archambault14}. A detection with a statistical significance of $71\sigma$ was made from the data on the night of the flare, with a time-averaged integral flux above 200~GeV of $(2.24 \pm 0.06) \times 10^{-6} \;\text{photon} \;\text{m}^{-2}\; \text{s}^{-1}$ (or $\sim0.95$~C.~U.).

%%%%%%%%%
\subsubsection{VHE gamma-ray flux variability and the modelling of the flare profile}
\label{subsubsec:VHELC}
%%%%%%%%%
\begin{figure}[ht!]
%\plotone{BLLac_20161005_LC200GeV_VERITAS_MCMC2models.pdf}
\hspace{-1cm}
\fig{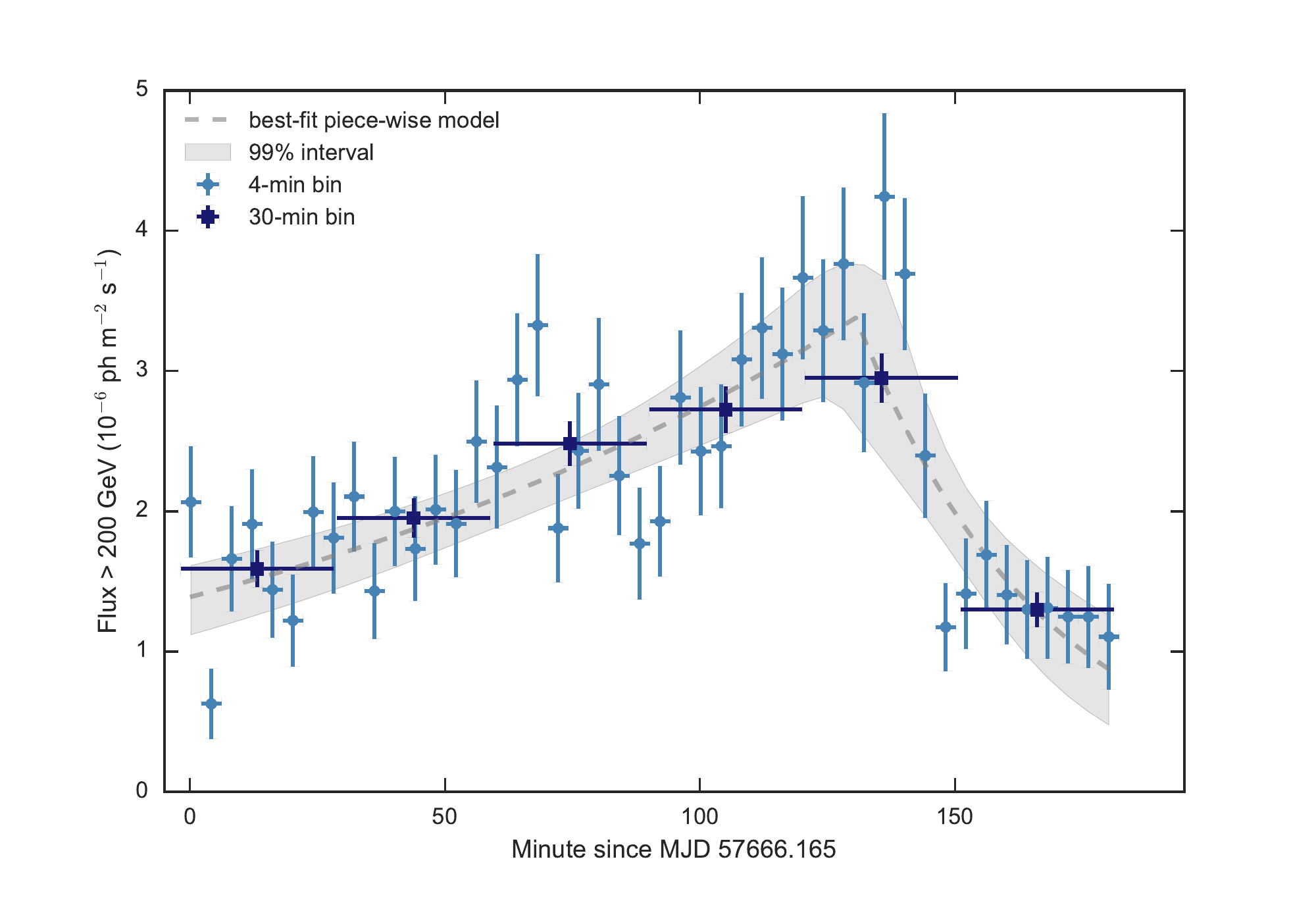}{0.58\textwidth}{}
\caption{The VERITAS TeV gamma-ray light curves of BL~Lacertae above $200$ GeV on 2016 Oct 5 (minute zero corresponds to 03:57:36 UTC). The light blue filled circles and the dark blue squares show the light curve in 4-min and 30-min bins, respectively. The grey dashed line shows the model (see Equation~\ref{flare_func}) with the best-fit parameters and the shaded region illustrates the 99\% confidence interval, both of which are derived from simulations using Markov chain Monte Carlo sampling. 
\label{fig:vlc}}
\end{figure}
Figure~\ref{fig:vlc} shows the VERITAS TeV gamma-ray light curve of BL~Lacertae above 200 GeV on 2016 Oct 5 with 4-minute and 30-minute bins. %A constant fit to the 4-minute binned light curve yields a p-value of $1.1 \times 10^{-16}$ ($\chi^2$=170.8 for 45 degrees of freedom; DOF), rejecting the hypothesis of steady flux. 
A gradual rise of the TeV flux by a factor of $\sim$2 followed by a faster decay was observed. 
The measured peak flux for the 30-minute-binned light curve is $(3.0 \pm 0.2) \times 10^{-6} \;\text{photon} \;\text{m}^{-2}\; \text{s}^{-1}$, corresponding to $\sim1.25$~C.~U., and that for the 4-minute-binned light curve is $(4.2 \pm 0.6) \times 10^{-6} \;\text{photon} \;\text{m}^{-2}\; \text{s}^{-1}$, or $\sim1.8$~C.~U.

We first fitted the 4-minute-binned VERITAS light curve with a constant-flux model, obtaining a $\chi^2$ value of 170.8 for 45 degrees of freedom (DOF), corresponding to a $p$-value of $1.1\times10^{-16}$ and rejecting the constant-flux hypothesis. 

To quantify the rise and decay times of the TeV flare, we then fitted the VHE gamma-ray light curve with a piecewise exponential function as follows: 
\begin{equation}
\label{flare_func}
F(t) = 
\begin{cases}
%F_{\text{peak}} e^{(t - t_{\text{peak}})/t_{\text{rise}}}, & t \leqslant t_{\text{peak}};  \\
%F_{\text{peak}} e^{-(t- t_{\text{peak}})/t_{\text{decay}}}, & t \geqslant t_{\text{peak}}; 
F_0 e^{(t - t_{\text{peak}})/t_{\text{rise}}}, & t \leqslant t_{\text{peak}};  \\
F_0 e^{-(t- t_{\text{peak}})/t_{\text{decay}}}, & t > t_{\text{peak}}; 
\end{cases} 
\end{equation}
where %$F_{\text{peak}}$ 
$F_0$ is the peak flux, $t_{\text{peak}}$ is the time of the peak flux, and $t_{\text{rise}}$ and $t_{\text{decay}}$ are the rise and decay times, respectively, on which the flux varies by a factor of $e$. 
%Flare profile \citep{Sato08}: 

The optimal values of the parameters and their uncertainties were determined from the posterior distributions obtained from Markov chain Monte Carlo (MCMC) simulations, for which the Python package \texttt{emcee}~\citep{Foreman-Mackey13} was used. 
The MCMC chain contains 100 random walkers in the parameter space initialized with a uniform random prior. Each random walker walks 4000 steps, the first 2000 steps of which are discarded as the ``burn-in'' samples. This amounts to $2\times10^{5}$ effective MCMC simulations. 
%The random walk step size is scaled
A proposal scale parameter was chosen so that the mean proposal acceptance fraction is 37\%, ensuring an adequate yet efficient sampling of the posterior distributions. 
Note that the parameters are bounded to be positive, so that they are physically meaningful, and sufficiently large upper bounds were also provided for computational efficiency. 
After the posterior distributions were obtained, kernel density estimation with Gaussian kernels of bandwidths equal to 1\% of the range of the corresponding parameter was used to estimate the most likely value (maximum a posteriori) and the 68\% confidence interval of each parameter. 
\begin{figure}[ht!]
%\plotone{MCMC_exp_model_tp200_a3p5.pdf}
\fig{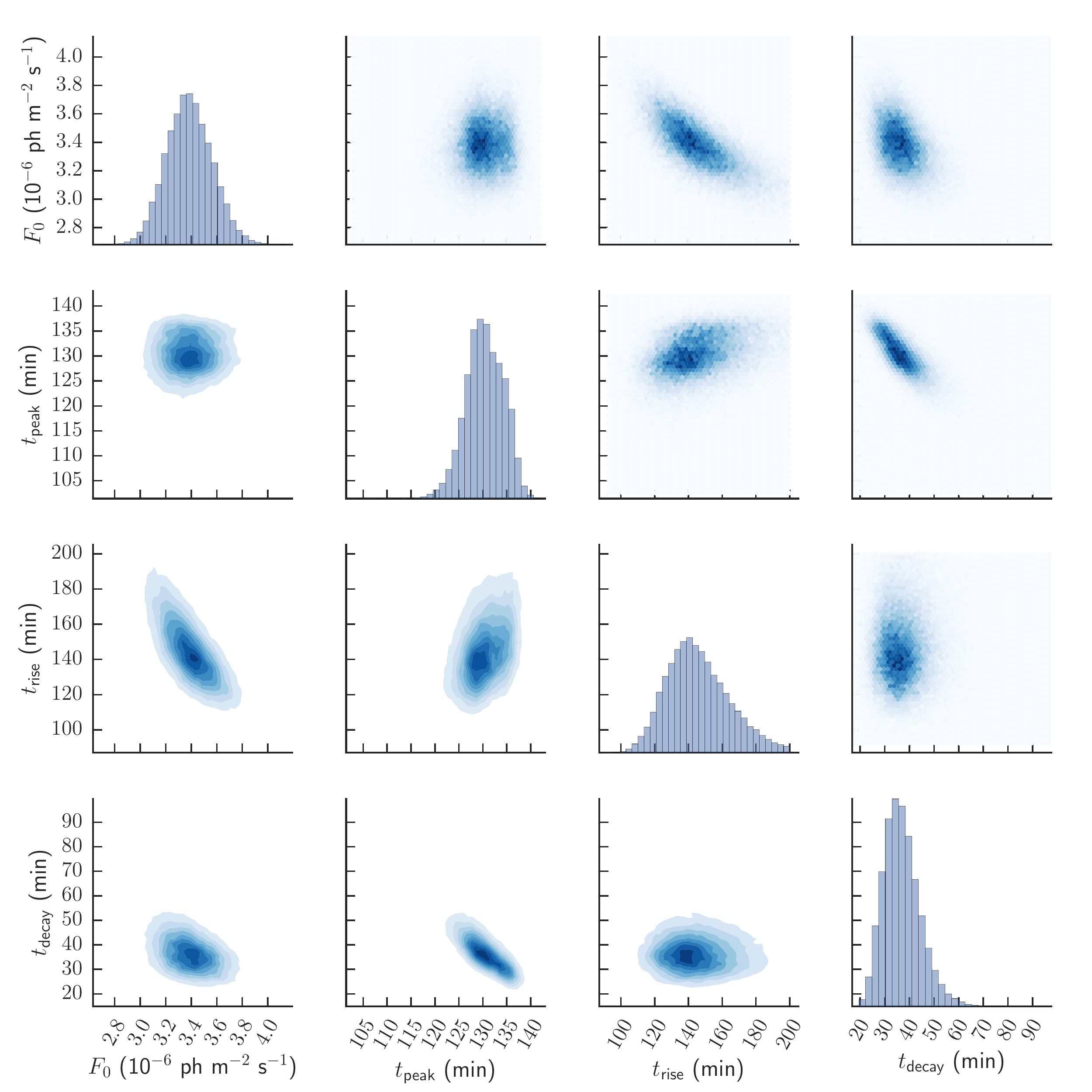}{0.49\textwidth}{}
\caption{The joint posterior distributions of the parameters in Equation~\ref{flare_func} obtained via MCMC simulations. The diagonal plots show the probability distribution histograms of individual parameters; the upper and lower diagonal plots show the two-dimensional histograms with hexagon binning and the kernel density estimations of the joint posterior distributions, respectively. 
\label{fig:mcmc_exp}}
\end{figure}

The joint posterior distributions of the parameters from the MCMC sampling are shown in Figure~\ref{fig:mcmc_exp}. 
The diagonal plots show the posterior probability distributions of each parameter, some of which (e.g., $t_{\text{peak}}$) appear non-Gaussian. 
Correlations between $t_{\text{peak}}$ and $t_{\text{rise}}$, as well as between $t_{\text{peak}}$ and $t_{\text{decay}}$, are also apparent in the off-diagonal joint distributions. 
The best-fit model and the 99\% confidence intervals from the MCMC sampling are shown in Figure~\ref{fig:vlc}. 
The rise and decay times of the flare are determined to be $140^{+25}_{-11}$ min and $36^{+8}_{-7}$ min, respectively. The best-fit peak time and flux are $130^{+5}_{-3}$ min (after MJD 57666.165) and $3.4^{+0.2}_{-0.2} \times 10^{-6}$ photon m$^{-2}$ s$^{-1}$, respectively.

Further VERITAS observations of BL~Lacertae were made on 2016 Oct 6 with 37.6-minute live exposure, and from Oct 22 to Nov 19 with 294.6-minute live exposure, after data-quality selection; neither of these sets of observations led to a detection of the source (signal significances of only $2.6\sigma$ and $0.9\sigma$, respectively). 
The integral flux upper limits (shown in the first panel of Figure~\ref{fig:mLC1}) between 0.2 and 30 TeV at 99\% confidence level from the observations on Oct 6 and between Oct 22 and Nov 19 were obtained as $2.0\times10^{-7}  \;\text{photon} \;\text{m}^{-2}\; \text{s}^{-1}$ and $2.8\times10^{-8}  \;\text{photon} \;\text{m}^{-2}\; \text{s}^{-1}$, respectively, assuming a power-law spectrum with a photon index of $3.3$ (see Section~\ref{subsec:VHESpec}). 

Motivated by the existence of multiple radio emission zones identified in VLBA data (see~Section~\ref{sec:VLBA}) and several multi-zone models for BL~Lacertae that are consistent with past observations \citep[e.g.,][]{Raiteri13, Hervet16}, we also fitted the light curve with a model including an additional constant-flux baseline. 
In a multi-zone model, different zones can be of different sizes and vary independently on different timescales. Therefore, it is possible to have a larger emitting zone that varies slowly which can be adequately described by a constant-baseline component on the timescale considered, and a smaller, more energetic zone that is responsible for the rapid flare described by the exponential components. 
With the more complex model, the best-fit decay time is only $2.6^{+6.7}_{-0.8}$ min, with a baseline flux of $1.2^{+0.1}_{-0.2}\times10^{-6} \;\text{photon}\; \text{m}^{-2} \text{s}^{-1}$. This best-fit baseline flux is higher than the upper limit obtained from the observations on the next day, 
indicating that the slower component would be required to vary on timescales of $\sim$1 day, 
%indicating the slower component varies on the timescale of $\sim$1 day, 
consistent with the GeV gamma-ray observations (Section~\ref{subsec:LAT}). However, we would like to highlight that it is not possible to unambiguously reject either model based on the statistics. 

%However, regardless of the models and fitting procedures used, the decay timescale of the VHE flare is smaller than an hour. 

%%%%%%%%%
\subsubsection{The VHE spectrum}
\label{subsec:VHESpec}
%%%%%%%%%
A power-law fit to the VERITAS spectrum of BL~Lacertae %yields a reduced $\chi^2/\text{DOF}$ value of 34 
yields a best-fit photon index of $3.28$ and a reduced $\chi^2$ value $\chi^2/\text{DOF}=30.6$, indicating that a simple power law does not adequately describe the spectrum. 
%and the best-fit parameters as follows: 
%\begin{equation}
%\label{eqVspecPL}
%\begin{split}
%\frac{dN}{dE}=& (1.03\pm 0.06)\times 10^{-7} \\ & 
%\times \left( \frac{E}{1\;\text{TeV}} \right)^{(-3.28 \pm 0.04)} %\\ & 
%\text{m}^{-2}\; \text{s}^{-1}\; \text{TeV}^{-1}.
%\end{split}
%\end{equation}

A log-parabola model %with a fixed pivot energy of 0.2 TeV 
fits the VERITAS spectrum better: 
\begin{equation}
\label{eqVspecLP}
\begin{split}
\frac{dN}{dE}&= (2.22\pm 0.07)\times 10^{-5} \\ & 
\times \left( \frac{E}{0.2\;\text{TeV}} \right)^{\left[ -(2.4 \pm 0.1) -  ( 1.8 \pm 0.3 ) \log_{10} (\frac{E}{0.2\;\text{TeV}})  \right]} \\ & 
\text{m}^{-2} \; \text{s}^{-1} \; \text{TeV}^{-1}, 
\end{split}
\end{equation}
with $\chi^2/\text{DOF}=1.6$. 

After de-absorbing the VHE spectrum using the optical depths for a source at a redshift of 0.069 according to the extragalactic background light model in \citet{Dominguez11}, the best-fit log-parabola model becomes: 
\begin{equation}
\label{eqVspecLP2}
\begin{split}
\frac{dN}{dE}&= (2.36\pm 0.07)\times 10^{-5} \\ & 
\times \left( \frac{E}{0.2\;\text{TeV}} \right)^{\left[ -(2.2 \pm 0.1) -  ( 1.4 \pm 0.3 ) \log_{10} (\frac{E}{0.2\;\text{TeV}})  \right]} \\ & 
\text{m}^{-2} \; \text{s}^{-1} \; \text{TeV}^{-1}, 
\end{split}
\end{equation}
with $\chi^2/\text{DOF}=1.7$. 
The observed and de-absorbed TeV gamma-ray spectra are shown together with the GeV gamma-ray spectra (Section~\ref{subsec:LAT}) in Figure~\ref{fig:gammaSED} in the $\nu F_\nu$ representation.

%%%%%%%%%
\subsection{\textit{Fermi}-LAT}
\label{subsec:LAT}
%\subsection{High-energy gamma-ray observations with Fermi-LAT}
%%%%%%%%%
The Large Area Telescope (LAT) on board the \textit{Fermi} satellite is a pair-conversion gamma-ray telescope sensitive to energies from $\sim$20~MeV to $>$300~GeV \citep{Atwood09}. 
%With the recently developed event-level analysis (Pass 8) by the Fermi-LAT collaboration \citep{Atwood13} 

An unbinned likelihood analysis was performed with the LAT ScienceTools \texttt{v10r0p5} and Pass-8 \texttt{P8R2\_SOURCE\_V6\_v06} instrument response functions \citep{Atwood13}. \texttt{SOURCE} class events with energy between 100 MeV and 300 GeV within 10$^\circ$ from the position of BL~Lacertae were selected. 
For the short durations of interest to the TeV flare, a simple model containing BL~Lacertae, another point source 3FGL~J2151.6+4154 $\sim2^\circ$ away from BL~Lacertae, and the contributions from the Galactic (\texttt{gll\_iem\_v06}) and isotropic (\texttt{iso\_P8R2\_SOURCE\_V6\_v06}) diffuse emission were included. A maximum zenith angle cut of 90$^\circ$ was applied. We checked in the residual test-statistics map that no significant excess was left unaccounted for within the model. For the short durations, a power law was used to model BL~Lacertae instead of the log-parabola model used in the \texttt{3FGL} catalog. %for the short durations. 
We verified with an analysis using a log-parabola spectral model and obtained consistent flux values.

For the light curve shown in the second panel of Figure~\ref{fig:mLC1}, an unbinned likelihood analysis was performed on each one-day interval, leaving the normalizations and power-law indices of BL~Lacertae and 3FGL~J2151.6+4154 free, as well as the normalization of the diffuse components. The source was in an elevated GeV gamma-ray state when the TeV flare was observed, although the GeV flux varied on a much longer timescale. %although on a much longer timescale. 
An exponential fit to a 15-day interval around the TeV gamma-ray flare yields a rise time of $2.1\pm0.2$ days and a decay time of $7\pm2$ days. 

\begin{figure}[ht!]
 \centering 
 \leavevmode 
 \includegraphics[width=1.0\linewidth]{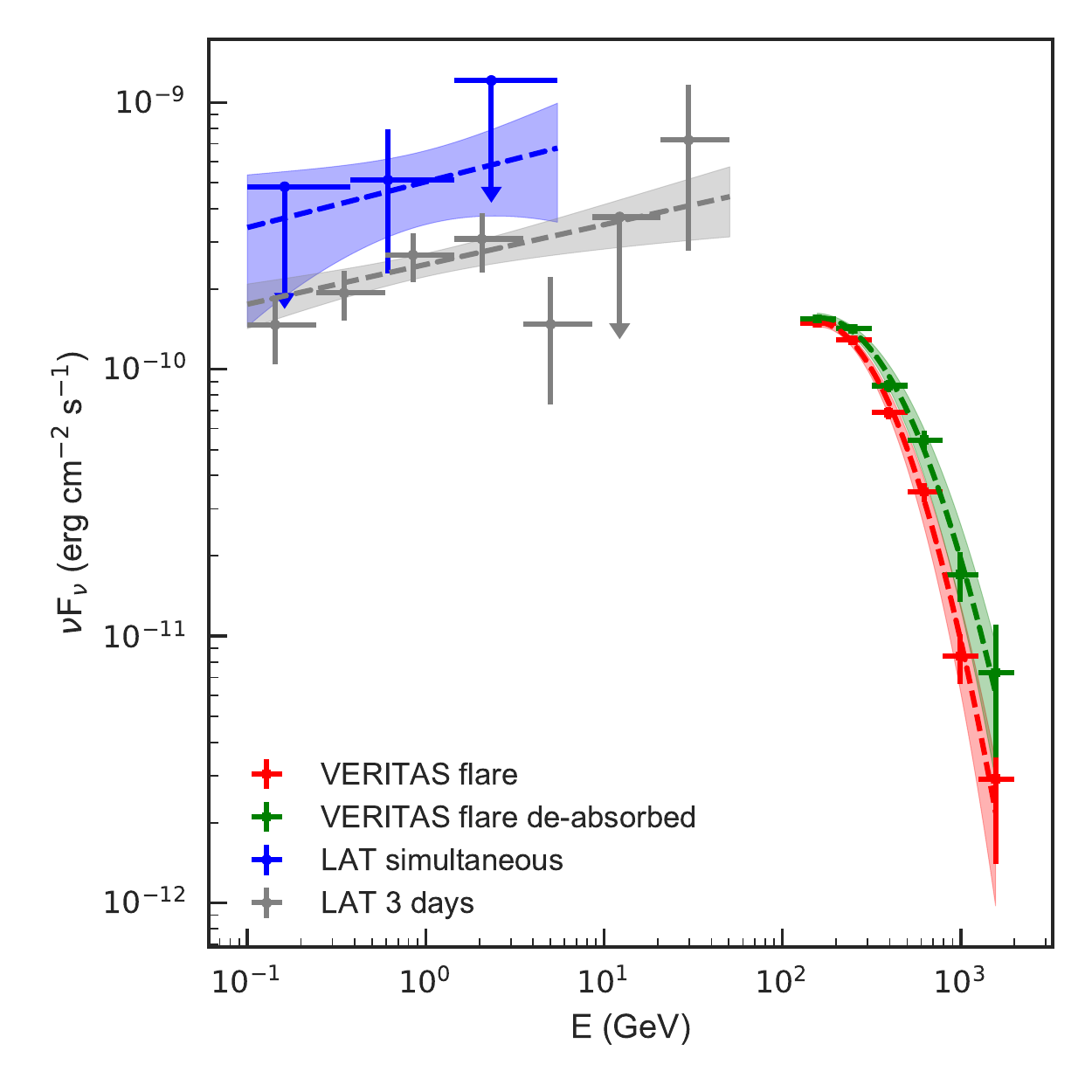}
 \caption{The gamma-ray SEDs of BL~Lacertae measured by {\it Fermi}-LAT and VERITAS. The {\it Fermi}-LAT SEDs strictly simultaneous with VERITAS observations on the night of the flare (2016 Oct 5) and from the three days around it are shown in blue and grey, respectively. The observed and de-absorbed VERITAS SEDs averaged over all observations on the night of the flare are shown in red and green, respectively. Each shaded region is derived from the 1-$\sigma$ confidence intervals of the best-fit parameters for the corresponding spectrum. %power-law model from the LAT unbinned likelihood analysis. 
 \label{fig:gammaSED}}
\end{figure}

The gamma-ray SEDs measured by the {\it Fermi}-LAT and VERITAS on the night of the TeV flare are shown in Figure~\ref{fig:gammaSED}. 
In order to obtain the GeV gamma-ray SEDs, we used the user-contributed tool \texttt{likeSED.py}\footnote{https://fermi.gsfc.nasa.gov/ssc/data/analysis/user/} to perform the unbinned likelihood analysis in several energy bands. 
%The best-fit power-law indices of BL~Lacertae for the strictly simultaneous and the three-day LAT data are $1.83\pm0.21$ and $1.85\pm0.07$, respectively. 
The power-law index that gives the best fit to the LAT data completely simultaneous with VERITAS is $1.83\pm0.21$, which is similar to that of the three-day binned LAT data, $1.85\pm0.07$. These values indicate a harder GeV gamma-ray spectrum during the flare than that reported in the \texttt{3FGL} catalog \citep[with an index of $2.25$;][]{Acero15}. 

Both the GeV and TeV gamma-ray spectral indices of this flare in 2016 are comparable to those of the flare in 2011 \citep{Arlen13}, 
and they suggest that the peak energy of the gamma-ray SED during the flare is between 5~GeV and 100~GeV. 

%%%%%%%%%
\subsection{\textit{Swift} XRT}
\label{subsec:XRT}
%%%%%%%%%
The X-Ray Telescope (XRT) on board the \textit{Swift} satellite is a grazing-incidence focusing X-ray telescope, and is sensitive to photons in the energy range 0.2--10~keV \citep{Gehrels04, Burrows05}. 

Follow-up observations of BL~Lacertae were carried out with the \textit{Swift}-XRT on 
%Swift followed up on BL~Lacertae on 
2016 Oct 6, 7, and 8; the only other XRT observations within a 45-day window around the time of the VHE flare were made on 2016 Oct 27 and Nov 2. 
The XRT data, taken in the photon-counting (PC) mode, were analyzed using the \texttt{HEAsoft} package (v6.19). % and \texttt{XSpec}. 
%The event files are calibrated and cleaned using the calibration files from 2011 September 5. The data were taken in the photon counting (PC) mode, and were selected from grades 0 to 12 over the energy range 0.3-10~keV. 
The data were first processed using \texttt{xrtpipeline} (v0.13.2) with calibration database (\texttt{CALDB v20160706}). 
%Pile-up effect was checked for all observations 
The count rates in the PC mode were $>0.5$~count~s$^{-1}$, and the effect of potential pile-up was checked for all observations by fitting a King function to the point-spread functions (PSFs) at $>$15~arcsecond \citep{Moretti05}. Those central pixels where the data fall below the model curve, indicating pile-up, were excluded. 

For the observations on 2016 Oct 6, the King function agrees with the data even on the brightest pixels. Therefore, a circular source region of a radius of 20 pixels centered on BL~Lacertae was used. 
For the data taken on 2016 Oct 7 and 8, annular source regions were used, with inner radii of four and two pixels, and an outer radius of 20 pixels. 
For all three observations, an annular background region with inner and outer radii of 70 and 120 pixels, respectively, was used. 
Note that source regions excluding the central two and four pixels were also tested for the observations on 2016 Oct 6, and consistent results were obtained. Therefore, we are confident that no bias was introduced by the different exclusion regions used for pile-up correction. 

The observations on Oct 7 consisted of two intervals of duration 486 s and 1422 s, separated by roughly one satellite orbital period ($\sim$90 min). A sustained dark stripe (likely due to bad CCD columns) appears in the XRT image near the position of BL~Lacertae, contaminating the second interval. Therefore, we conservatively chose to use only the data recorded during the first interval. The image and spectrum of each $\sim$3 min of this relatively short exposure were checked for data quality, and no anomaly was found. 

\begin{deluxetable}{ccccc}[h]
\tablecolumns{5}

\tablecaption{\textit{Swift}-XRT spectral-fit results using the absorbed-power-law model described in Equation~\ref{eqXspec}, with $N_H$ free and fixed. The errors quoted denote 68\% confidence intervals. 
\label{tab:Xspec}}
%\tablewidth{700pt}
\tabletypesize{\scriptsize}
\tablehead{
\colhead{Date} & \colhead{$\alpha$} & 
\colhead{K} & \colhead{$N_H$} & $\chi^{2}/\text{DOF}$ \\ 
\colhead{} & \colhead{}  & 
\colhead{$10^{-2}$ keV$^{-1}$cm$^{-2}$s$^{-1}$} & \colhead{$10^{21}$ cm$^{-2}$}  & \colhead{}
} 
\startdata
       Oct 6		& $ 2.5 \pm 0.1 $ 	&  $ 0.62^{+0.07}_{-0.06} $ 	& $ 2.7^{+0.3}_{-0.3} $	& 0.83 	\\
       Oct 7 		& $ 2.1 \pm 0.1 $	&  $ 4.6^{+0.6}_{-0.5} $ 	& $ 3.1^{+0.5}_{-0.4} $	& 1.07 	\\
       Oct 8 		& $ 2.3 \pm 0.1 $	&  $ 0.43^{+0.06}_{-0.05} $	& $ 3.0^{+0.5}_{-0.4} $	& 0.54 	\\        
       Oct 27 	& $ 1.4^{+0.4}_{-0.3} $	&  $ 0.14^{+0.08}_{-0.04} $	& $ 1.2^{+2.2}_{-1.2} $	& 0.48 	\\ 
       Nov 2		& $ 1.3 \pm 0.3 $	&  $ 0.11^{+0.04}_{-0.03} $	& $ 1.4^{+1.1}_{-0.8} $	& 0.33 	\\ \hline
       Oct 6		& $ 2.54 \pm 0.07 $ 	&  $ 0.65 \pm 0.03 $ 		& $ 2.9 $ (fixed)& 0.82 	\\
       Oct 7 		& $ 2.08 \pm 0.07 $	&  $ 4.34  \pm 0.24 $ 	& $ 2.9 $ (fixed)& 1.04	\\
       Oct 8 		& $ 2.26 \pm 0.08 $	&  $ 0.41 \pm 0.02  $		& $ 2.9 $ (fixed)& 0.52 	\\   %\hline
       Oct 27 	& $ 1.64 \pm 0.20 $	&  $ 0.19  \pm 0.03 $ 	& $ 2.9 $ (fixed)& 0.50	\\
       Nov 2 	& $ 1.76 \pm 0.19 $	&  $ 0.17  \pm 0.02 $ 	& $ 2.9 $ (fixed)& 0.63	\\
       %Oct 6		& $ 2.20 \pm 0.05 $ 	&  $ 0.47 \pm 0.02 $ 	& $ 1.83 $	 (fixed)& 1.03 	\\
       %Oct 7 		& $ 1.80 \pm 0.06 $	&  $ 3.14  \pm 0.16 $ 	& $ 1.83 $	 (fixed)& 1.49	\\
       %Oct 8 		& $ 1.95 \pm 0.07 $	&  $ 0.30 \pm 0.02  $	& $ 1.83 $	 (fixed)& 0.94 	\\
\enddata
\end{deluxetable}

\begin{figure}[ht!]
 \centering 
 \leavevmode 
  \includegraphics[width=1.0\linewidth]{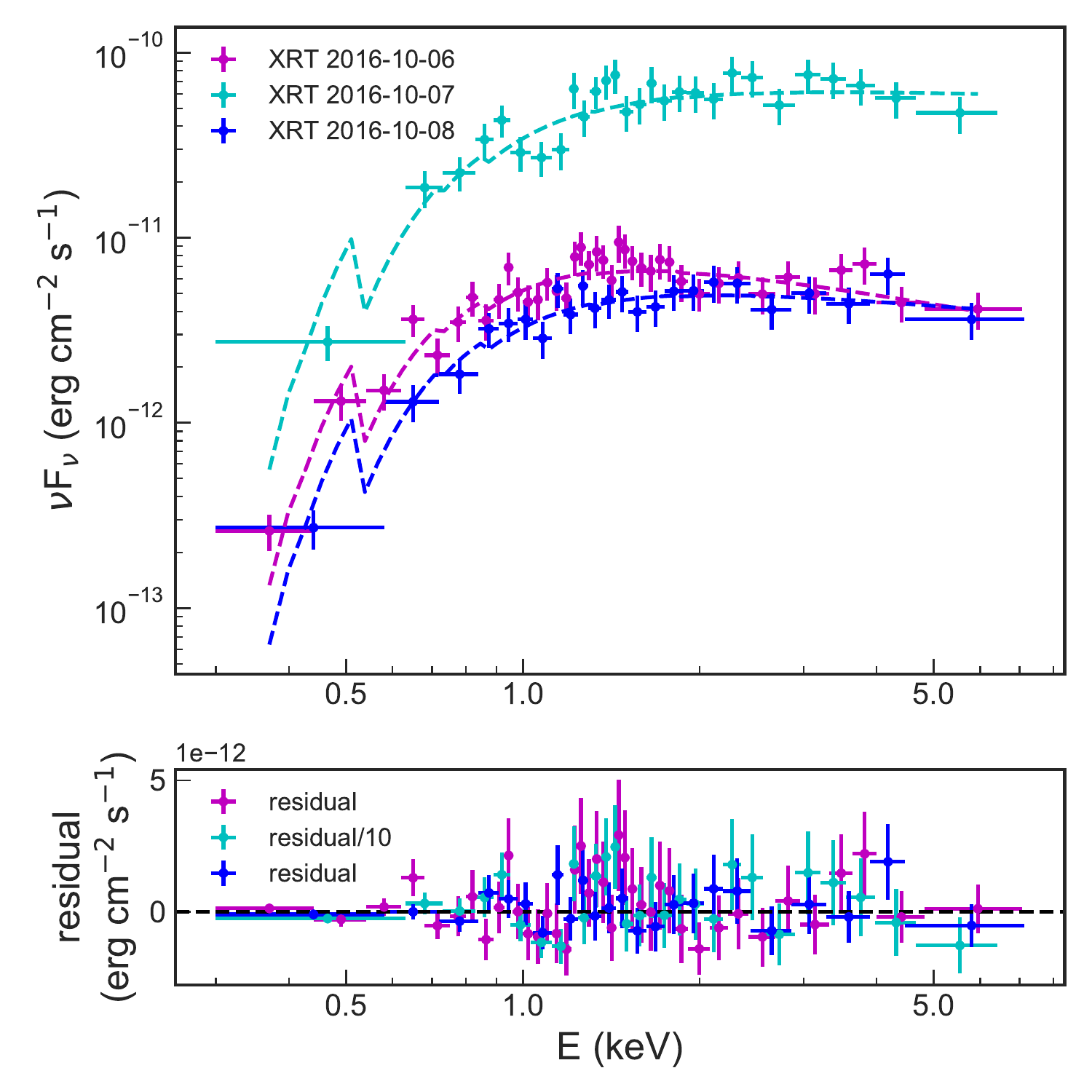}
 \caption{{\it Top}: the X-ray SEDs measured by {\it Swift}-XRT on 2016 Oct 6, 7, and 8. The dashed lines are the best-fit absorbed-power-law model with $N_H$ fixed at $2.9  \times10^{21} \;\text{cm}^{-2}$. {\it Bottom}: the distributions of the fit residuals of each X-ray SED. Note that the residual values on Oct 7, shown in cyan, are divided by 10 to facilitate comparison with the other two distributions shown. 
 \label{fig:XSED}}
\end{figure}

\begin{figure*}[ht!]
%\plotone{f5.pdf}
\includegraphics[width=1.0\linewidth]{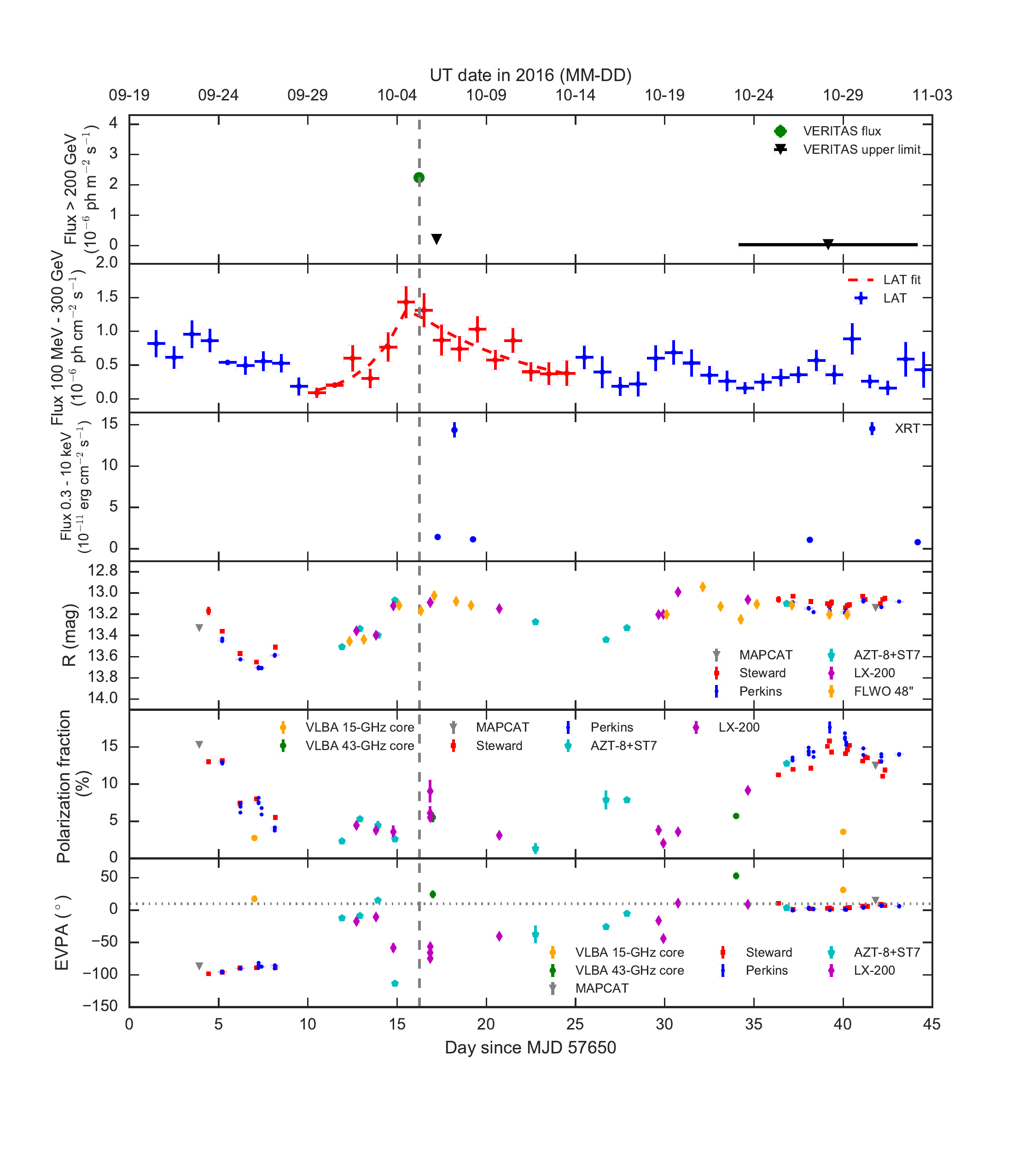}
\vspace{-4.0em}
\caption{The 45-day MWL light curves of BL~Lacertae around the time of the VHE flare. The top panel shows the TeV gamma-ray flux measured by VERITAS on the night of the flare, as well as the upper limits obtained later. The second panel shows the daily-binned GeV gamma-ray light curve measured by {\it Fermi}-LAT, as well as a piecewise exponential fit (see Equation~\ref{flare_func}) to a 15-day interval around the TeV gamma-ray flare (red dashed line). The third panel shows the X-ray energy flux from the five {\it Swift}-XRT observations. The bottom three panels show the $R$-band photometric and polarimetric measurements (see~Section~\ref{sec:opt}). The grey vertical dashed line shows the peak time of the TeV gamma-ray flare observed by VERITAS. The fractional polarizations and the EVPAs of the 43-GHz and 15-GHz core are also shown in the bottom two panels. The grey horizontal dotted line in the bottom panel shows the position angle of the jet. 
\label{fig:mLC1}}
\end{figure*}

Ancillary response files were generated using the \texttt{xrtmkarf} task with the response matrix file \texttt{swxpc0to12s6\_20130101v014.rmf}. 
The spectrum was fitted with an absorbed-power-law model (\texttt{po*wabs}): 
\begin{equation}
\label{eqXspec}
\frac{dN}{dE} = e^{-N_H\sigma(E)} K \left( \frac{E}{1\;\text{keV}} \right)^{-\alpha}, 
\end{equation}
where $N_H$ is the column density of neutral hydrogen, $\sigma(E)$ is the photoelectric cross-section, and $K$ and $\alpha$ are the normalization and index of the power-law component, respectively. 
The best-fit values of the parameters are shown in Table~\ref{tab:Xspec}. Note that the best-fit values of $N_H$ are in agreement with the archival results from X-ray spectral fit \citep{Madejski99,Ravasio03,Arlen13}, %\citet{Arlen13} and \citet{Ravasio03}, 
but are larger than the value $N_H = 1.8  \times10^{21} \;\text{cm}^{-2}$ from the Leiden/Argentine/Bonn (LAB) survey of Galactic HI \citep{Kalberla05}. This difference is likely due to the additional contribution of the Galactic molecular gas (e.g., CO emission), as BL Lacertae is relatively close to the Galactic plane (with a Galactic latitude $b=-10.44^\circ$) \citep{Madejski99}. 
%\citep{Madejski99} N-H
%W
As the $N_H$ value is expected to stay constant over the period of interest, we also fit the same model with $N_H$ fixed at %the value from the LAB survey, as well as 
the average best-fit value $N_H = 2.9  \times10^{21} \;\text{cm}^{-2}$ over the three nights of XRT observations, to better constrain the spectral index and normalization. %we fix it at $N_H = 2.9  \times10^{21} \;\text{cm}^{-2}$. 
We also investigated an absorbed-log-parabola model with $N_H$ fixed at $2.9 \times10^{21} \;\text{cm}^{-2}$ to fit the X-ray spectra. We see no evidence for spectral curvature, as the best-fit log-parabola model reduces to a power law.
The X-ray SEDs of BL~Lacertae measured on 2016 Oct 6, 7, and 8 are shown in Figure~\ref{fig:XSED}. 
The X-ray emission from the source was strongest and hardest on 2016 Oct 7 (two days after the TeV gamma-ray flare) compared to the day before and the day after (see Table~\ref{tab:Xspec}). 
The energy flux values based on the best-fit absorbed-power-law model between 0.3~keV and 10~keV on 2016 Oct 6, 7, and 8 were $(1.4\pm 0.1)$, $(14.2\pm 0.9)$, and $(1.1\pm 0.1)$ $\times10^{-11}$ erg cm$^{-2}$ s$^{-1}$, respectively, as shown in the third panel of Figure~\ref{fig:mLC1} along with the results from two more observations taken on 2016 Oct 27 and Nov 2. Note that the

%%%%%%%%%
\subsection{Optical facilities}
\label{sec:opt}
%\subsection{$R$-band polarization}
%%%%%%%%%
%BL~Lacertae was monitored in R band at a high cadence by a number of optical instruments, including the Steward Observatory, AZT-8+ST7, Perkins, LX-200, and MAPCAT. 
BL~Lacertae was monitored in the $R$-band at high cadence by a number of optical facilities, including the Steward Observatory\footnote{\url{http://james.as.arizona.edu/~psmith/Fermi}} \citep{Smith09}, the AZT-8 reflector of the Crimean Astrophysical Observatory, the Perkins telescope\footnote{\url{http://www.lowell.edu/research/research-facilities/1-8-meter-perkins/}}, the LX-200 telescope in St. Petersburg, Russia, 
%and the Monitoring AGN with Polarimetry at the Calar Alto Telescopes (MAPCAT)\footnote{\url{http://www.iaa.es/~iagudo/research/MAPCAT/MAPCAT.html}} program. 
and the Calar Alto 2.2-m Telescope (with observations obtained through the MAPCAT\footnote{MAPCAT stands for the Monitoring of AGN with Polarimetry at the Calar Alto Telescopes, see: \url{http://www.iaa.es/~iagudo/_iagudo/MAPCAT.html}} program) in Almer\'ia, Spain. 
%We also included the $r'$-band observations made with the 48-inch telescope at the Fred Lawrence Whipple Observatory (FLWO). We transformed the $r'$-band flux to $R$-band using the color index $V-R=0.7$ for BL Lacertae \citep{Fan98} and the empirical transformation $r-R=(0.267\pm0.005)(V-R)$ \citep{Jordi06}. Note that we do not consider any variability of the color index of the source. 
We also included the $r'$-band observations made with the 48-inch telescope at the Fred Lawrence Whipple Observatory (FLWO). We transformed the $r'$-band flux to $R$-band  using the color index $V-R=0.73 \pm 0.19$ for BL~Lacertae \citep{Fan98} and the %empirical transformation $r-R = (0.267 \pm 0.005)(V-R) + (0.088 \pm 0.003)$ \citep{Jordi06}. %The errors from the $r'$ photometry are shown in Figure~\ref{fig:mLC1}. 
transformation $r'-R = 0.19 (V-R) + 0.13$ \citep{Smith02}. 
The dominant uncertainty from the conversion to $R$-band comes from the variability of the $V-R$ color index, resulting in an additional systematic uncertainty of $\sim$0.04 magnitude, which is included in the converted FLWO $R$-band magnitude shown in Figure~\ref{fig:mLC1}. %If the color index change during the epoch of observations shown, it would be an overall zero-point shift of all the FLWO points.
Any variability in the color index $\Delta(V-R)<0.19$ during the epoch of observations shown would lead to a shift of the FLWO $R$-band magnitude within the error bars shown. 

The $R$-band flux and polarization measurements contemporaneous with the gamma-ray flare are shown in Figure~\ref{fig:mLC1}. 
The lower three panels (from top to bottom) show the $R$-band magnitude, polarization fraction, and electric vector position angle (EVPA) of the source, respectively. 
%The fifth panel shows the difference between each EVPA and the previous one. 
A $-180^\circ$ shift is applied to all the EVPA measurements before MJD 57662, so that the EVPA difference between MJD 57662 and 57658 is reduced to $\sim$80$^\circ$ from $\sim$100$^\circ$ before the shift was applied \citep[see e.g.,][]{Abdo10PolRot}. 
%The measurements from all instruments agree with one another.  
The measurements are reasonably consistent between the various instruments. 

%The grey dashed line shows the peak time of the TeV flare observed by VERITAS. 
The $R$-band flux from the source varied in a similar manner to the GeV flux, with an increase  
observed a few days before the VHE flare. 
The optical EVPA appeared to have rotated smoothly from roughly perpendicular to the position angle (PA) of the jet in late 2016 Sept to roughly parallel in late 2016 Oct, except for three days before the TeV gamma-ray flare when the optical EVPA was nearly aligned with the PA of the jet. This was followed by a sudden decrease in EVPA on the day before the TeV gamma-ray flare. 
The fractional polarization was relatively low around the time of the TeV flare, and increased to the highest value of the 45-day period in late Oct when the EVPA was again aligned with the jet. 
%
%Such behaviour is potentially consistent with the emergence of a radio knot observed by VLBA described below (in Section~\ref{sec:discuss}), 
%as it may add a component with a different EVPA to the observed total flux and cancel the net polarization. 

\begin{figure*}[ht!]
%\hspace{-1cm}
\fig{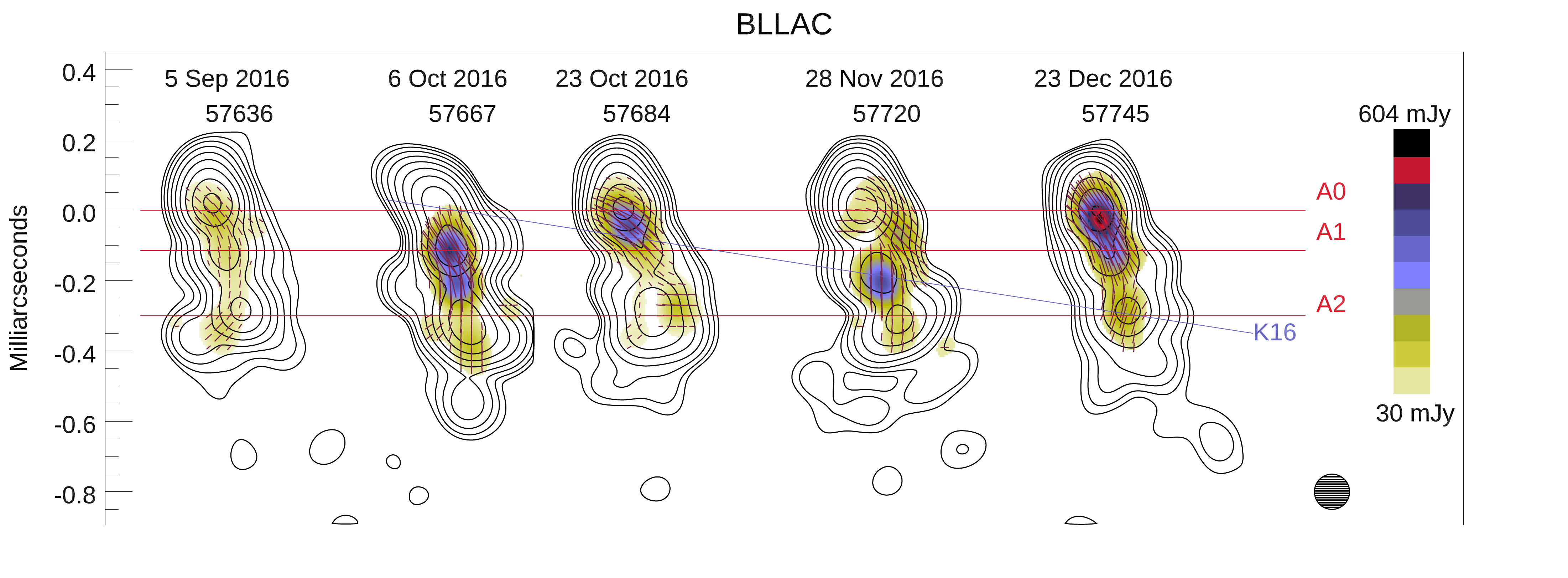}{1.05\textwidth}{}
%\fig{BLLAC_vlba_fall2016.eps}{0.58\textwidth}{}
\caption{
The 43 GHz VLBA total (contours) and polarized (color scale) intensity images
of BL Lac. The total intensity peak is 1.15 Jy/beam. The contours are 0.2, 0.4, 0.8,\ldots, 51.2, 96\% of the peak. The restoring beam shown in the right bottom corner
is a circular Gaussian with FWHM=0.1~mas. Linear segments within the images indicate
position angles of the polarization, with the length of segments proportional to the local polarized intensity. Red horizontal lines mark the mean locations of the three quasi-stationary features; the blue line across the epochs from 2016 October to December traces the motion of the superluminal knot $K16$.  
%Total (shown as contours) and linearly polarized (shown as colored map) intensity images of BL~Lacertae from VLBA observations at 43 GHz. The images are convolved with a circular Gaussian restoring beam of FWHM = 0.1 mas. Tick marks are separated by 0.05 mas. The contours are in multiples of $\sqrt{2}$ times the peak of 1.15 Jy/beam. Red horizontal lines indicate the mean centroid locations of the three quasi-stationary components, $A0$, $A1$, and $A2$, while the white line shows the motion of moving knot $K16$.
\label{fig:VLBA}}
\end{figure*}

%%%%%%%%%
%\subsection{VLBA}
\subsection{Radio facilities}
%%%%%%%%%
\label{sec:VLBA}
%BL~Lacertae was regularly observed with VLBA at 43~GHz and 86~GHz as part of the monitoring program of gamma-ray bright blazars at Boston University \footnote{\url{http://www.bu.edu/blazars/VLBAproject.html}}. 
%A series of 43-GHz VLBA images of BL~Lacertae were taken on 2016 Jul 31, Sep 5, Oct 6, Oct 23, and Nov XX. 
%86~GHz ~ 3.5 mm; 43~GHz ~ 7 mm

%The data were correlated at the National Radio Astronomy Observatory in Socorro, NM, and analyzed at Boston University following the procedures described by \citet{Jorstad05}. 
%No significant change in the structure of the inner jet was observed from Jul 31 to Sep 5, 
%while changes between Sep 5 and Oct 6, as well as between Oct 6 and Oct 23, were observed. 

%...(more after VLBA data are delivered, core shift as a function of frequency + RM)

BL~Lacertae was observed throughout the period of interest at 43 GHz with the Very Long Baseline Array under the VLBA-BU-BLAZAR monitoring program \citep{Jorstad16} and at 15.4 GHz with the Monitoring Of Jets in Active galactic nuclei with VLBA Experiments (MOJAVE) program \citep{Lister09}. The 43-GHz and 15.4-GHz VLBA data calibration and imaging procedures were identical to those described by \citet{Jorstad05} and \citet{Lister09}, respectively. 

%BL~Lacertae is also in the sample of the Monitoring Of Jets in Active galactic nuclei with VLBA Experiments (MOJAVE) program. For this work, we only used results from polarization measurements at 15.4 GHz. The data reduction procedures are described by \citet{Lister09}. Briefly, the flux density of the core component is derived from a Gaussian model fit to the interferometric visibility data. Polarization properties of the core are then derived by taking the mean Stokes Q and U flux densities of the nine contiguous pixels that are centered at the Gaussian peak pixel position of the core fit. The results include fractional linear polarization, electric vector position angle (note the 180~degree degeneracy), and polarized flux densities. The flux density has an uncertainty of $\sim 5\%$, and the position angle of polarization has an uncertainty of $\sim 3$~degrees.

Figure~\ref{fig:VLBA} presents 43-GHz images of the parsec-scale jet of BL~Lacertae at five epochs from 2016 Sept 5 to Dec 23. The second epoch, 2016 Oct 6, took place only one day after the VHE flare. The images are convolved with a circular Gaussian restoring beam with a full width at half maximum (FWHM) of 0.1 mas, % milliarcsec (mas), 
which is similar to the angular resolution of the longest baselines along 
the (southern) direction of the jet. 
We note that the Oct 6 observation was affected by equipment failure at the Mauna Kea and Hancock antennas, at the extremities of the array, although this degraded the north-south angular resolution by only 14\%. The corresponding linear resolution at the redshift of BL~Lacertae is 0.13 pc in projection on the sky and $1.8^{+0.8}_{-0.4}$ pc if we adopt a viewing angle of $4.2^\circ\pm1.3^\circ$ between the jet axis and line of sight \citep{Jorstad17}. 

As was the case in previous observations \citep{Jorstad05, Arlen13, Gomez16, Wehrle16}, the main structure of the jet consists of three quasi-stationary brightness peaks, designated as $A0$, $A1$ 0.12 mas to the south of $A0$, and $A2$ 0.30 mas to the south of $A0$. The locations of $A1$ and $A2$ appear to fluctuate as moving emission features (frequently referred to as ``knots'') with superluminal apparent velocities pass through the region. Such combination of moving and stationary emission components complicates the interpretation of the changing structures of the total and polarized intensities. Because of this, the interpretation that we offer to explain the variations within the images is not unique. 

We ignore the effects of Faraday rotation on the polarization EVPA, which \citet{Jorstad07} estimated to be low ($-16^\circ$) between 43~GHz and 300~GHz. 
It is worth mentioning that \citet{Hovatta12} measured a much lower (by an order of magnitude) Faraday rotation using 8-GHz to 15-GHz observations. 
This could be due to a combination of a possible variability in the rotation measure and a decrease of the rotation measure with distance from the central black hole \citep{Jorstad07}, as the core at 15~GHz is located further away from the black hole compared to the core at 43~GHz due to the effect of opacity. 

A knot of emission with enhanced polarization at 43 GHz, which we designate as $K16$, appears to propagate down the jet. 
%It moves by 0.23 mas between Oct 23, when its centroid is $\sim$0.05 mas south of $A0$, and Dec 23, when its centroid is $\sim$0.28 mas from $A0$. 
Its centroid moves from $\sim$0.05 mas south of $A0$ on Oct 23 to $\sim$0.28 mas from $A0$ on Dec 23. 
This corresponds to an apparent speed of $6c$, within the range typically observed in BL~Lacertae \citep{Jorstad05, Marscher08, Arlen13, Lister13, Wehrle16, Jorstad17}. Extrapolation back to Oct 6 places the knot $K16$ 0.01 mas north of the centroid of $A0$, within the $A0$ emission region characterized by its angular size of $0.03\pm0.02$ mas \citep{Jorstad17}. This implies that the VHE flare occurred as the moving knot crossed the stationary ``core'', which \citet{Marscher08} have interpreted as a standing shock located $\sim$1 pc from the central black hole.

%\begin{figure}[ht!]
\begin{figure*}[ht!]
\hspace{-1cm}
\gridline{\fig{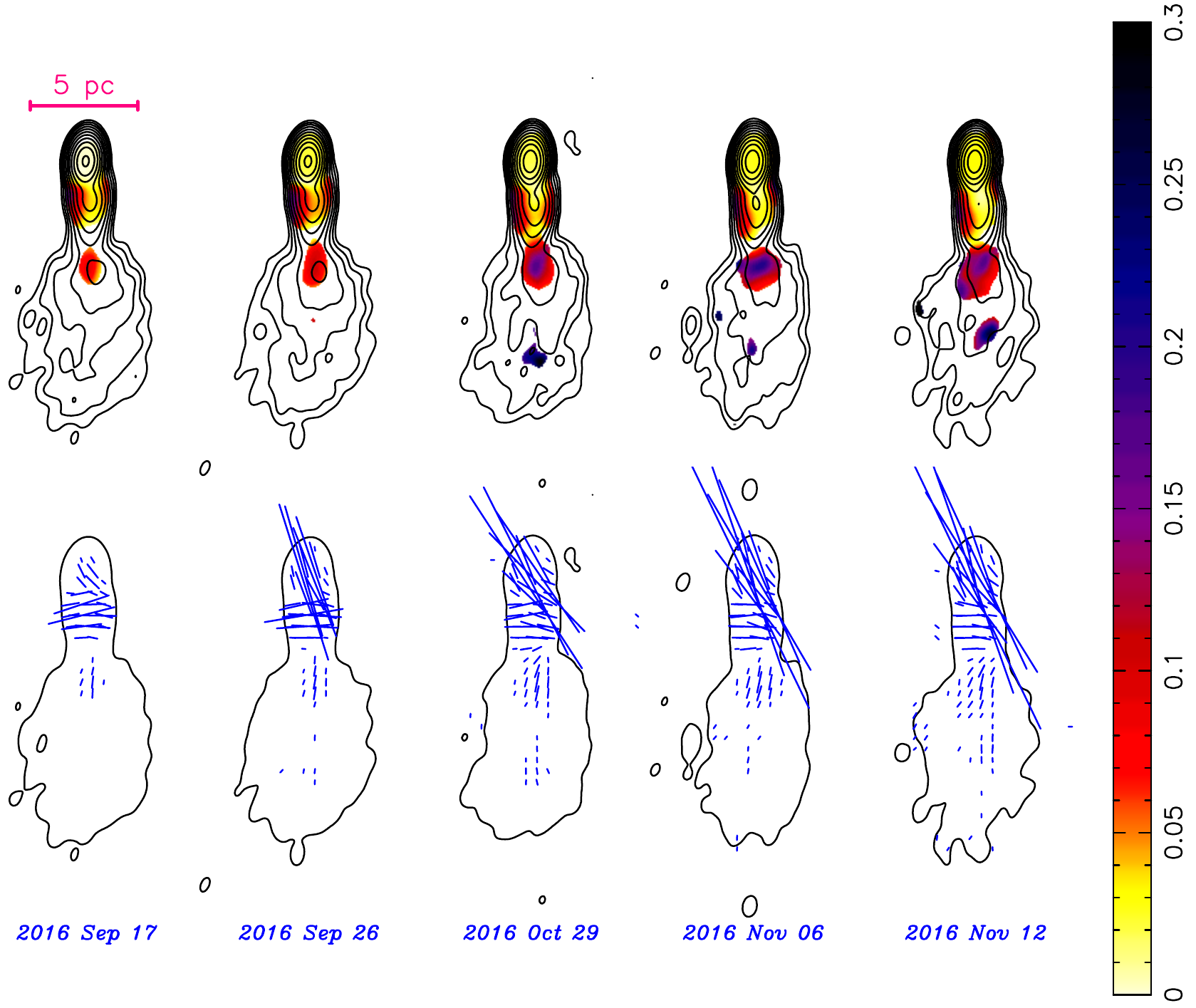}{0.65\textwidth}{}}
\gridline{\fig{{f7b}.pdf}{0.65\textwidth}{}}
\vspace{-1em}
\caption{Images of BL~Lacertae from VLBA observations at 15.4 GHz for ten epochs. A Gaussian restoring beam with dimensions 0.883 mas $\times$ 0.56 mas and a position angle $-8.2^\circ$ was used. 
The colors in the top rows of each panel show the fractional polarized level. The direction of the blue line segments in the bottom rows illustrate the EVPA, and their length corresponds to polarized intensity, the lowest of which shown is 0.5 mJy/beam. 
The contours show the total intensity, with a base contour of 1.1 mJy/beam in both top and bottom rows, and successive contours increment by factors of two in the top rows. 
%plotted every 8th pixel, The pixel scale is 0.05 mas/pixel
The typical total and polarized intensity image rms values in these images are 0.09 mJy/beam and 0.1 mJy/beam, respectively. 
%The angular scale of the image is 1.29 pc/mas. The polarization color scale in each image is shown on the side. 
\label{fig:VLBA15}}
\end{figure*}
%\end{figure}

The VLBA images at 15.4~GHz, as shown in Figure~\ref{fig:VLBA15}, reveal the evolution of the jet structures %further away from the central source on a larger spatial scale compared to 43 GHz, as a result of optical depth and angular resolution, respectively. 
further away from the central source and on a larger spatial scale, as a result of optical depth and angular resolution, respectively, compared with the observations at 43 GHz. 
Therefore, a delay is expected between the measurements at these two frequencies. 
The polarized intensity of the stationary core of BL~Lacertae at 15.4 GHz reached a minimum on 2016 Dec 26 and gradually increased, %with a bright feature appearing at $\sim$1 mas southwest of the core, which may correspond to the knot $K16$ observed at 43 GHz earlier. 
with a potentially bright feature with distinct polarization angle (consistent with the EVPAs measured at 43 GHz on 2016 Dec 23), %on 2016 Dec 10 and 26, 
which may correspond to the knot $K16$ observed at 43 GHz earlier, appearing at $\sim$1 mas southwest of the core. 
This is consistent with past observations of the same source with the VLBA at different frequencies reported by \citet{Bach06}, where new components of the jet were seen to fade as they separated from the core, disappearing at  $\sim$0.7~mas %and 1 mas, %between about 0.7 mas and 1 mas, 
and reappearing at $\sim$1~mas. 

%\citep{D'arcangelo09}
We show the fractional polarizations and EVPAs of the 43-GHz and 15-GHz core in the bottom two panels of Figure 5, along with the $R$-band results. 
The EVPAs of the core at 43 GHz and 15 GHz are roughly consistent with the PA of 10$^\circ$ of the jet over the course of a few months since 2016 Sept. This implies that the magnetic field is toroidal/helical at the core, as we discuss in Section~\ref{sec:discuss}. 
Further downstream in the jet, the EVPAs become more perpendicular to the PA of the jet, as shown in Figure~\ref{fig:VLBA15}. Such location-dependent radio EVPAs help us to interpret the dominant optical component based on the optical EVPA data. 

\begin{figure}[ht!]
%\plotone{BLLac_longterm_radio_LC.pdf}
\fig{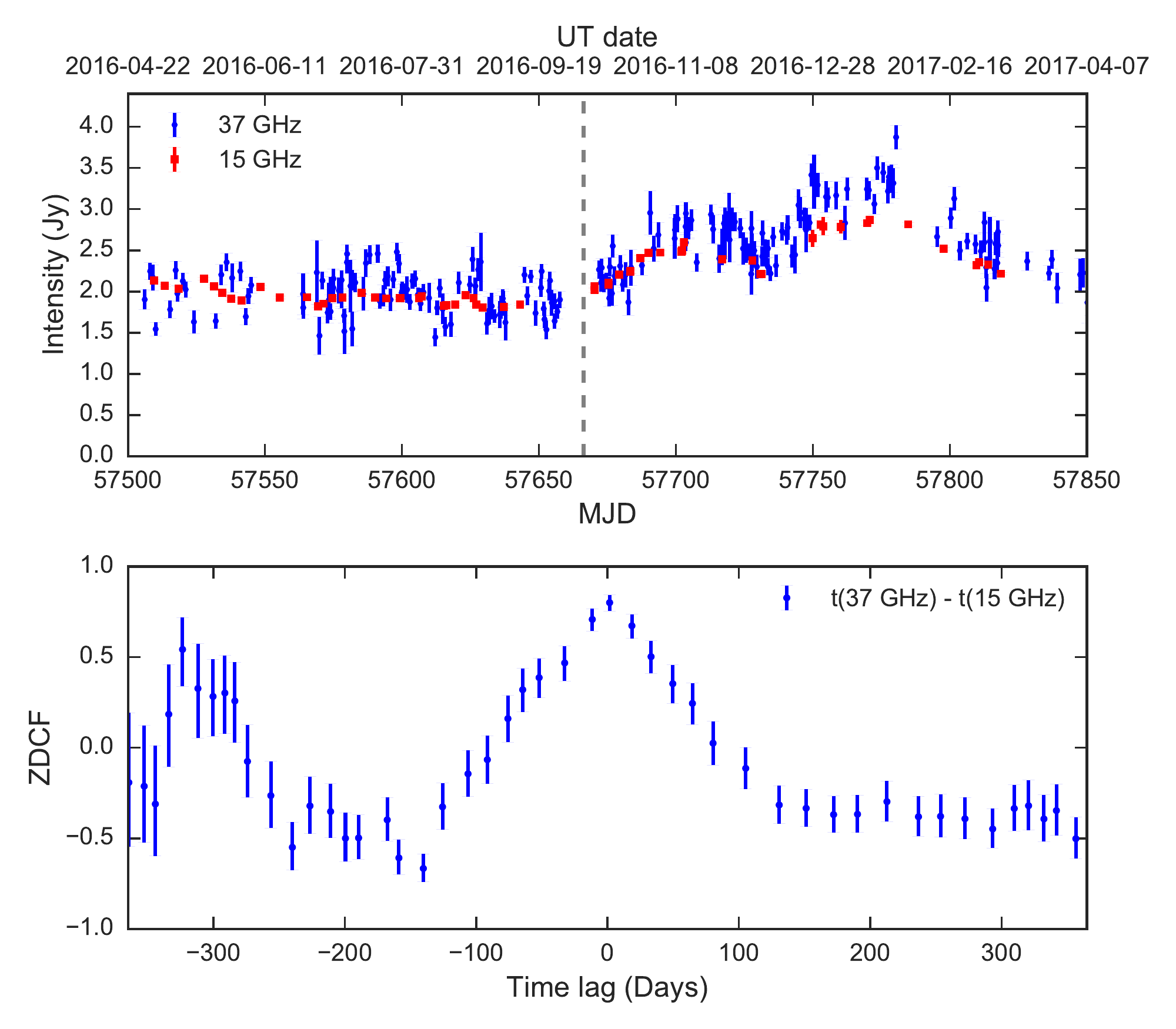}{0.5\textwidth}{}
\caption{The top panel shows the 37-GHz (blue dots) and 15-GHz (red squares) radio light curves measured over $\sim1$ yr by Mets\"ahovi and OVRO, respectively. The grey dashed line shows the peak time of the TeV flare observed by VERITAS. The bottom panel shows the $z$-transformed discrete cross correlation between the two light curves above. The time lag values are calculated as the difference in time $t$ between 37~GHz and 15~GHz so that positive time lags correspond to the 37-GHz flux leading the 15-GHz flux. %No significant time lags or leads are observed. 
\label{fig:radio_corr}}
\end{figure}
We show in Figure~\ref{fig:radio_corr} the evolution of the total flux density of BL~Lacertae, measured by the Mets\"ahovi Radio Observatory (MRO) at 37~GHz and by Owens Valley Radio Observatory (OVRO) at 15~GHz, respectively, over about a year, as well as their z-transformed discrete cross-correlation \citep[ZDCF;][]{Alexander13}. 

The 37-GHz observations were made with the 13.7-m diameter Aalto
University Mets\"ahovi radio telescope, which is a radome-enclosed
Cassegrain-type antenna situated in Finland. The measurements were made with a 1-GHz-bandwidth dual-beam
receiver centered at 36.8 GHz. The %HEMPT (
high electron mobility pseudomorphic transistor front end operates at room temperature. %The
%observations are Dicke switched ON--ON observations, alternating the
%source and the sky in each feed horn. 
The 37-GHz observations are Dicke-switched ON--ON observations, alternating the
source and the sky in each feed horn to remove atmospheric and ground contamination. 
Typical integration time to
obtain one flux-density data point is between 1200 s and 1600 s. The
detection limit of the telescope at 37 GHz is on the order of 0.2 Jy
under optimal conditions. Data points with a signal-to-noise ratio $<$ 4
are handled as non-detections.

The flux-density scale is set by observations of the HII region of DR 21, while NGC 7027, 3C 274 and 3C 84 are used as secondary
calibrators. A detailed description of the data reduction and analysis
can be found in \citet{Terasranta98}. The error estimate in
the flux density includes the contributions from the measurement rms
and the uncertainty of the absolute calibration.

%The OVRO 40-m telescope is equipped with off-axis dual-beam optics and a cryogenic high electron mobility transistor low-noise amplifier with center frequency 15 GHz and bandwidth 3 GHz. 
The OVRO 40-m telescope uses off-axis dual-beam optics and a cryogenic pseudo-correlation receiver with a 
15.0-GHz center frequency and 3-GHz bandwidth. 
The source is alternated between the two beams in an ON--ON fashion to remove atmospheric and ground 
contamination. The fast gain variations are corrected using a 180$^\circ$ phase switch. 
%The two sky beams are Dicke switched using the off-source beam as a reference, and the source is alternated between the two beams in an ON-ON fashion to remove atmospheric and ground contamination. 
Calibration is achieved using a temperature-stable diode noise source to compensate for receiver gain drifts, and the flux-density scale is derived from observations of 3C~286 assuming the value of 3.44~Jy at 15.0~GHz reported by \citet{Baars77}. The systematic uncertainty in the flux density scale is $\sim$5\%, which is not included in the error bars in Figure~\ref{fig:radio_corr}. Complete details of the reduction and calibration procedures are given in \citet{Richards11}.

The 37-GHz and 15-GHz light curves show that at the time of the TeV gamma-ray flare, BL~Lacertae was transitioning from a steady radio flux state to a flaring state that lasted for about five months. 
The ZDCF shows no significant detection of any time lag between the fluxes at the two frequencies, suggesting that both observations are dominated by the flux from a region that is optically thin at 15 GHz. 
%We note that the increase of the ZDCF coefficient around $-$300 days is likely due to the finite length of the observations. 

%%%%%%%%%%%%%%%%%%
%%%%%%%%%%%%%%%%%%
%%%% %     Discussion     %%%%
%%%%%%%%%%%%%%%%%%
%%%%%%%%%%%%%%%%%%
\section{Discussion}
\label{sec:discuss}
%A fast TeV gamma-ray flare was observed from BL~Lacertae for the second time in VHE by VERITAS. % with 
%For the second time at VHE, a fast gamma-ray flare from BL~Lacertae has been observed by VERITAS. 
For the second time, VERITAS has detected a fast gamma-ray flare from BL~Lacertae. 
%For the second time by VERITAS, a fast gamma-ray flare from BL~Lacertae has been observed. 

%In contrast, the VERITAS measurements during the previous VHE fast flare from the object did not provide information about the rising phase. 
While no information was obtained from the rising phase of the first VHE flare in 2011 \citep{Arlen13}, the VERITAS measurements during the 2016 flare described in this work cover both the rise and decay phases of the flare. 
%Duty cycle of the jet shooting blobs? 
%Are these two flares the same or different? GeV, X-ray?

\subsection{On the size of the gamma-ray-emitting region}
The fastest timescale of a flare (in this case the decay time) provides a constraint on the size $R$ of the emitting region, as %$R\le\text{c}t_\text{decay} \delta /(1+z)$, 
\begin{equation}
\label{eqR}
R\le \frac{\text{c}t_\text{decay} \delta} {1+z}, 
\end{equation}
where $c$ is the speed of light, $\delta$ is the Doppler factor of the jet\footnote{$\delta = [\Gamma (1-\beta \cos \theta)]^{-1}$, where
$\Gamma$ is the bulk Lorentz factor of the jet, and $\theta$ is the angle between the axis of the jet and the line of sight.}, and $z$ is the redshift of the source. 

The mass of the central black hole ($M_\text{BH}$) of BL~Lacertae was estimated to be $\sim3.8\times 10^8 M_\odot$ by \citet{Wu09} using the $R$-band absolute magnitude and the empirical correlation between black hole mass and bulge luminosity of the host galaxy \citep{McLure02}. %\citep{Xie04}, \citep{Gupta12} 
%Using the above value of $M_\text{BH}$, 
The corresponding Schwarzschild radius $R_s$ of the central black hole of BL~Lacertae is $\sim1.1\times 10^{12}$~m ($\sim 3.6\times 10^{-5}$~pc). 
%3.09e-08 arcsec
It is worth noting that the mass measurement of a black hole is a challenging task, and $M_\text{BH}$ values in the range $(0.16 - 5.01)\times 10^8 M_\odot$ have been reported for BL~Lacertae \citep[see][and references therein]{Gupta12}. 

The Doppler factor $\delta$ was estimated to be $\sim$24 according to the method described in \citet{Hervet16} from the propagation of a possible perturbation in the radio jet observed with VLBA at 15 GHz~\citep{Lister13}, assuming a viewing angle of $2.2^\circ$ based on radio apparent-velocity measurements. Taking the best-fit value of $t_\text{decay}=36^{+8}_{-7}$ min (see Section~\ref{subsubsec:VHELC}) and using Equation~\ref{eqR}, we estimate the upper limit on the size of the emitting region to be $R \lesssim11.9 R_s$. %$R \le 11.9^{+2.6}_{-2.3} R_s$. 
%Similarly, the much shorter best-fit decay timescale in model II leads to a stronger constraint on the size of the emitting region: $R \le 0.9^{+2.2}_{-0.3} R_s$. 
% model I MCMC $ 11.9^{+2.6}_{-2.3} $ 
% t_decay = 1740 s (29 min), delta=23 => R<= 1.12e13m ~ 3.6e-4 pc ~ 9.6 R_Schwarzschild
% t_decay = 2160 s (36 min), delta=23 => R<= 1.39e+13m ~ 4.5e-4 pc ~ 11.9 R_Schwarzschild
% t_decay = 2640 s (44 min), delta=23 => R<= 1.7e13m ~ 5.5e-4 pc ~ 14.5 R_Schwarzschild
% model II MCMC $ 0.9^{+2.2}_{-0.3} $
% t_decay = 108 s (1.8 min), delta=23 => R<= 7e11m ~ 2.3e-5 pc ~ 0.6 R_Schwarzschild
% t_decay = 156 s (2.6 min), delta=23 => R<= 1.0e+12m ~ 3.3e-5 pc ~ 0.86 R_Schwarzschild
% t_decay = 558 s (9.3 min), delta=23 => R<= 3.6e12m ~ 11.7e-5 pc ~ 3.1 R_Schwarzschild
%Both model I and model II are justified, model I is motivated by the Occam's razor, and model II by multi-zone scenario. 
%
%
%calc_dL 0.069
%Luminosity distance: ouput values = 
%age at z, 	distance in Mpc, 	kpc/arcsec, 	apparent to abs mag conversion
%11.71 		271.47 			1.23 			37.31
% 1.23 pc/mas
%Note that model II is statistically preferred taking into account its better agreement with data, as well as the extra parameter (see Appendix). 

\subsection{On the gamma-ray flare profile}

%Slow rise and fast decay, what does it tell us? Sign of a brutal stop of the injection? 
An asymmetric profile with a faster decay of the VHE gamma-ray flux was observed in the flare, which would be caused by an abrupt cessation of the high-energy particle injection \citep[see e.g.,][]{Katarzynski03, Petropoulou16}. In this scenario, the flaring activity is attributed to fresh injection of high-energy particles into the emitting region instead of {\em in situ} acceleration of the particles. However, minimal variability in the radio band would be observable for this interpretation. 
Since strictly simultaneous radio observations were not performed, we cannot draw any conclusions regarding the radio variability at the time of the TeV gamma-ray flare. However, we note that the observed gamma-ray flare profile and the longer-term radio light curves (Figure~\ref{fig:radio_corr}) are consistent with the model proposed by \citet{Petropoulou16}. In this model, a fast gamma-ray flare can be produced by a small plasmoid in the magnetic reconnection layer, with no concurrent radio flares from the single plasmoid but a delayed radio flare powered by the entire reconnection event. The delay timescale is expected to correspond to the duration of the reconnection event, typically a few weeks. 
The asymmetric flare is in contrast to the more frequently observed flaring profile, a fast rise followed by a slow decay, which can be the manifestation of {\em in situ} acceleration and/or a longer cooling time (i.e., longer than the acceleration time) associated with a steep particle-energy distribution \citep[analogous to solar flares; see e.g.,][]{Harra16}. 

BL~Lacertae showed an enhancement in its GeV gamma-ray flux at the time of the TeV flare, but on a longer timescale of a few days. It also exhibited high X-ray flux on 2016 Oct 7 (two days after the TeV flare), %two days after the TeV flare on 2016 Oct 7, 
about a factor of 10 stronger than the flux on Oct 6 and 8. 
%These observations indicate strong activity of the relativistic particles in the jet. 
These observations indicate efficient acceleration of relativistic particles in the jet to at least a few hundred GeV. We note, however, that the delayed X-ray flare may or may not be related to the TeV gamma-ray flare, %since we cannot rule out the possibility that there was an X-ray flare simultaneous with the TeV gamma-ray flare due to the lack of strictly simultaneous X-ray data. 
since the lack of strictly simultaneous X-ray data precludes us from ruling out the possibility of an X-ray flare simultaneous with the TeV gamma-ray one. 
The different variability timescales of the observed TeV and GeV gamma rays give a hint that they may originate from different emitting zones. 
One possibility is that the GeV gamma rays were produced by particles injected into and accelerated in a large shock region \citep[e.g., a radio core; see][]{Kovalev09}, while the TeV gamma rays were produced through magnetic reconnection in a localized region \citep[e.g., a small plasmoid in a magnetic reconnection layer, possibly at the interface between a radio core and a moving knot; see][]{Petropoulou16}.

\begin{deluxetable*}{ccccccccccccccc}[ht]
\tablecolumns{15}
\tablecaption{ The parameters used to calculate the constraints shown in Figure~\ref{fig:constraints}. 
\label{tab:constraints}}
%\tablewidth{700pt}
\tabletypesize{\scriptsize}
\tablehead{
\colhead{$L_\text{gamma}$} & \colhead{$L_\text{syn}$} & \colhead{$L_\text{d}$\tablenotemark{a}} & 
\colhead{$t_\text{var}$} & \colhead{$M_\text{BH}$\tablenotemark{b}}  & \colhead{$E_\text{cool}$}  & 
\colhead{$\delta/\Gamma$}  & \colhead{$\epsilon_\text{BLR}$} &\colhead{$\epsilon_\text{IR}$} & 
\colhead{$r_\text{BLR}$} & \colhead{$r_\text{IR}$}  & 
\colhead{$E_\text{BLR}$} & \colhead{$E_\text{IR}$}  &
\colhead{$g_\text{SSC}$} & \colhead{$g_\text{ERC}$}  \\
\colhead{(erg s$^{-1}$)} & \colhead{(erg s$^{-1}$)} &  \colhead{(erg s$^{-1}$)} &  \colhead{(min)} &  \colhead{($M_\odot$)} & 
\colhead{(TeV)} &  \colhead{$$} &  \colhead{$$} &  \colhead{$$} &  \colhead{(pc)} & 
\colhead{(pc)} & \colhead{(eV)} &  \colhead{(eV)} &  \colhead{$$} &  \colhead{$$} 
} 
\startdata
    $7.8\times10^{45}$    &	$7.8\times10^{44}$	&  $6.0\times10^{44}$	&36	&  $3.8\times10^{8}$	&
    1	&1	&0.1	&0.1 	 &0.025 	&0.5	&10 	&0.3  	&0.75	&0.5	\\
\enddata
\tablecomments{$L_\text{gamma}$, $L_\text{syn}$, and $L_\text{d}$ are the observed gamma-ray luminosity, the synchrotron luminosity, and disk luminosity, respectively; $t_\text{var}$ is the observed variability time; $M_\text{BH}$ is the mass of the central black hole; $E_\text{cool}$ is the energy of the observed photons due to the external Compton cooling of relativistic electrons; $\delta/\Gamma$ is the ratio between the Doppler factor and Lorentz factor of the electrons; $\epsilon_\text{BLR}$, $r_\text{BLR}$, and $E_\text{BLR}$ are the covering factor, characteristic radius of the BLR, and the energy of BLR photons, respectively; $\epsilon_\text{IR}$, $r_\text{IR}$, and $E_\text{IR}$ are similar parameters for the IR-emitting torus region; $g_\text{SSC}$ and $g_\text{ERC}$ are the bolometric correction factors for SSC and ERC mechanisms. }
\tablenotetext{a}{\citet{Abdo11BLLac}}
\tablenotetext{b}{\citet{Wu09}}
\vspace{-2.5em}
\end{deluxetable*} 
%\vspace{-10em}

\subsection{On the Lorentz factor and the location of the gamma-ray-emitting region}

Without simultaneous MWL observations with temporal resolution comparable to that of the TeV gamma-ray observations, we cannot construct a reliable broadband SED of the source during the TeV flaring state. Instead, we constrain the Lorentz factor ($\Gamma$) of the gamma-ray-emitting region 
based on the gamma-ray variability, assuming two different emission mechanisms, SSC and ERC. 
Both models have been used to describe the broadband SED of BL~Lacertae in the past \citep[e.g.,][]{Madejski99, Raiteri13}. %, we note that the harder spectrum and the higher 
%\citet{Raiteri13} reached the conclusion 
However, we note that during the flare, the peak of the gamma-ray SED is located between $\sim5$ GeV and $\sim$100 GeV, higher than that in the lower flux state \citep[e.g.,][]{Abdo11BLLac, Rani13}. 
Such behaviour is most frequently observed in FSRQs and can be interpreted with ERC process on IR photons in the torus region \citep[e.g.,][]{Ghisellini09,Tagliaferri15}. Such ERC process was also used to interpret the emission of BL~Lacertae in a flaring state \citep{Madejski99, Ravasio03}. 

%During the flare, the Fermi-LAT spectrum presented a hard spectral index of ~1.8, corresponding to an high energy luminosity peak within the 10-100 GeV range when considering the VERITAS spectrum. This energy differs dramatically from the source low state when the high energy emission peaks typically before the Fermi-LAT energy range (~< 100 MeV). Such a flaring behavior is most of the time observed in FSRQs and interpreted as a strong dominance of Inverse-Compton processes of high energy particle on the nucleus thermal field. This scenario was applied with success to interpret the photon seed of an high energy bright past flare of BL Lac in 1997 (Madejski et al. 1999). The dramatic change of the brodband SED between quiet and flaring state was also studied by Ravasio et al. 2002. For the following we then consider that the high energy emission measured during the 2017 flare is strongly dominated by external inverse Compton process on the nucleus thermal field.

Assuming a one-zone SSC model, we can calculate an opacity constraint on the Doppler factor $\delta$ of the TeV gamma-ray emitting region by requiring the pair-production optical depth to be $\leqslant 1$, following Equation 3.7 and 3.8 in \citet{Dondi95} \citep[see also][]{Arlen13}. We found that $\delta \gtrsim 13$ using the following observables: the best-fit decay time of the TeV gamma-ray flare ($36$ min), the center of the highest-energy bin with significant excess of the TeV gamma-ray spectrum of the source during the flare ($\sim$1.5~TeV), the $R$-band magnitude inferred from the FLWO observations on the same night (13.17), and the near-infrared spectral index \citep[$1.5$;][]{Allen82}. Assuming a viewing angle of $2.2^\circ$, the constraint $\delta \gtrsim 13$ is equivalent to a constraint on the Lorentz factor of $\Gamma \gtrsim 7$. 

Assuming the gamma rays are emitted via an ERC process, we can constrain the Lorentz factor $\Gamma$ and the distance $r$ from the central black hole of the gamma-ray-emitting region
following the method described by \citet{Nalewajko14}. 
%This method can be reliable when the VHE zone is dominated by the EIC on thermal radiation field (i.e: close to the black hole)
The collimation constraint was derived from the requirement $\Gamma\theta\lesssim1$. 
Both SSC and ERC processes are considered in the calculation of the SSC constraint, while the majority of the gamma rays are assumed to be produced via ERC process. 
For the cooling constraint, only the ERC process on the thermal radiation fields close to the black hole is considered. 
%We do not take into account possible inverse-Compton process by an external synchrotron field, %which could be significant for this source.
This does not take into account any possible inverse-Compton scattering of an external synchrotron field, % (which we consider below), 
which, as we consider below, would loosen the cooling constraint on $\Gamma$ at large distances from the central black hole. 
We also assume that the emitting region is spherically symmetric. It is possible that the emitting region is not spherical (e.g., if it is passing a standing shock), and the constraints on $\Gamma$ and $r$ may change. 

The values of the parameters used for the calculation of the above three constraints on $\Gamma$ and $r$ are shown in Table~\ref{tab:constraints}. 
Some of the parameters are constrained by observations, and the other parameters are chosen so that a conservative constraint is derived. For example, we set the Compton dominance parameter $q=L_\text{gamma}/L_\text{syn}=10$ based on the observed $R$-band magnitude and the peak flux of the gamma-ray SED, the former of which should provide a good estimation of the peak of the synchrotron flux, considering that the source is a lower-frequency-peaked BL~Lac object. The SSC luminosity was set equal to the observed gamma-ray luminosity $L_\text{SSC}=L_\text{gamma}$ in order to obtain a conservative SSC constraint. We also used a relatively-high observed gamma-ray energy (1~TeV) for a conservative ERC cooling limit. 
% $R$-band mag 13.1 => 1e-14 ergs cm^-2 s^-1 A^-1 => R ~ 6580A => 6.6e-11 ergs cm^-2 s^-1
%
We note that 
changes in the values of the parameters describing the geometry of the external radiation fields, namely the covering factor $\epsilon_\text{BLR}$ and characteristic radius $r_\text{BLR}$ for the BLR, and similarly $\epsilon_\text{IR}$ and $r_\text{IR}$ for the IR torus, which are poorly constrained by observations, could change the cooling constraint. The values of the radii used in this work are derived based on the disk luminosity $L_\text{d}=6.0\times10^{44}$~erg~s$^{-1}$ \citep[][]{Abdo11BLLac} and the relations $r_\text{BLR} = 1\times 10^{15} \sqrt{(L_\text{d}/10^{45} \text{erg s}^{-1})}\;\text{m} \approx 0.025\; \text{pc}$ and $r_\text{IR} = 2\times 10^{16} \sqrt{(L_\text{d}/10^{45} \text{erg s}^{-1})}\;\text{m} \approx 0.5\; \text{pc}$ \citep{Ghisellini15}. 

\begin{figure}[ht!]
%\plotone{BLLac_longterm_radio_LC.pdf}
\fig{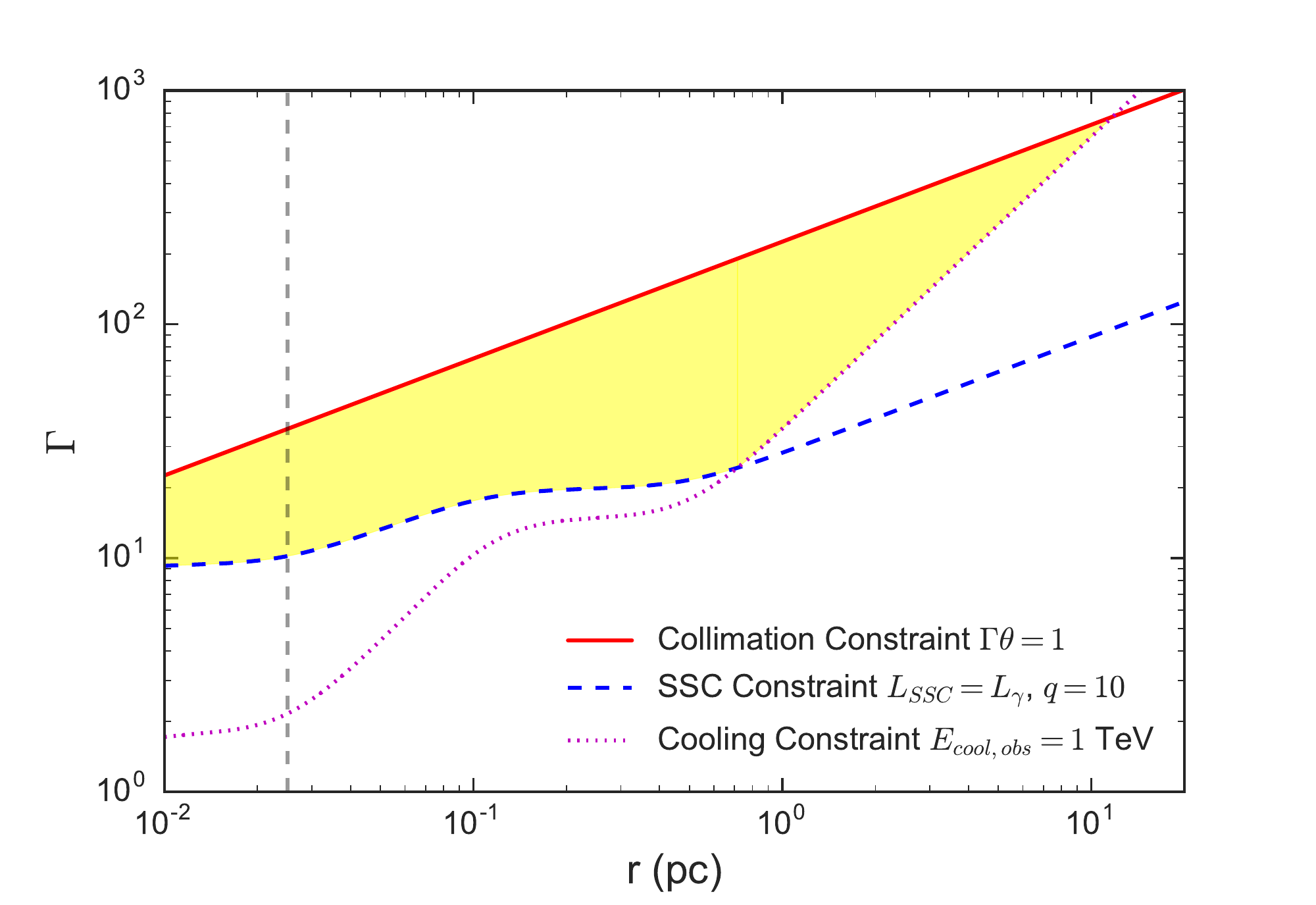}{0.51\textwidth}{}
\caption{The constraints on the Lorentz factor ($\Gamma$) and the distance ($r$) between the central black hole and the gamma-ray emitting location. The grey vertical dashed line indicates the location of the BLR (0.025 pc) used in the calculation. The yellow shaded region illustrates the allowed parameter space. 
\label{fig:constraints}}
\end{figure}

In this analysis, the distance $r$ between the central black hole and the gamma-ray-emitting region is constrained to be $\lesssim 12.4$ pc. 
If we fix the Lorentz factor at $\Gamma=24$, then we constrain the distance to be $0.01\lesssim r/\text{pc} \lesssim 0.7$. If we fix the distance at $r=1$ pc, the estimated distance between the core $A0$ and the central black hole, and assume that the gamma rays are produced as the knot $K16$ passes the core $A0$, then the Lorentz factor is only loosely constrained at $35 \lesssim \Gamma \lesssim 226$. 

At small $r$ values ($r\lesssim0.68$ pc), the SSC constraint on the lower limit of $\Gamma$ is stricter than the cooling constraint. At $r=r_\text{BLR}=0.025$ pc (the smallest distance for the VHE-emitting region without heavy absorption from the radiation field in the BLR), we put a strong lower limit on the Lorentz factor $\Gamma\gtrsim10.1$, which is larger than the archival values ($\sim$5--7) derived from radio observations \citep{Jorstad05,Jorstad17}, and consistent with the value of 24 adopted in this work. A possible explanation for the lower values of $\Gamma$ obtained from the radio observations is that they are calculated based on the apparent velocity of the superluminal features in the jet, which may travel at a lower speed compared to the bulk plasma flow \citep[e.g.,][]{Lister13, Hervet16}. 

At large $r$ values ($r\gtrsim2$ pc), the lower limit on the Lorentz factor $\Gamma$ increases to $>100$, exceeding the typical range of $\Gamma\sim4-50$ obtained from observations of blazars \citep[e.g.,][]{Jorstad05, Cohen07, Lister16}. This indicates that another seed-photon population, such as an external synchrotron radiation field, is needed if the gamma-ray-emitting region lies beyond $\sim$2 pc. 

\subsection{On the radio and optical polarizations}

The 43-GHz and 15-GHz observations reveal that the EVPAs at the core are mostly parallel to the PA of the jet. This implies that the magnetic field is likely toroidal or strongly helical near the core, consistent with earlier observations of BL Lacertae \citep[e.g.,][]{Gomez16}. The 15-GHz EVPAs at larger distances away from the core become more perpendicular to the PA of the jet, indicating that the magnetic field may be more poloidal in the outer jet. Such a magnetic field configuration has been proposed for low-frequency-peaked BL Lac objects \citep[e.g.,][]{Kharb08, Hervet16}. 

Based on these radio observations, we can use the observed changes in the optical polarizations of BL Lacertae (as shown in Figure~\ref{fig:mLC1}) to gain insights into the magnetic field structure and the location of the region that dominates the optical emission \citep[e.g.,][]{D'arcangelo09, Algaba11}. 
The optical EVPAs were observed roughly perpendicular to the PA of the jet in late 2016 Sept, indicating that the magnetic field is close to being aligned with the jet and likely dominated by the region downstream in the jet at that time. 
Similarly, the optical EVPAs became mostly parallel to the PA of the jet after late Oct, suggesting that the optical emission was then dominated by the core or the inner jet. 
We also observed the highest optical fractional polarization during this period, suggesting that %the core/inner jet is possibly more magnetized than the outer jet where magnetic energy is sufficiently dissipated. 
the magnetic field of the core/inner jet is more ordered. %within the dominant emission region. 

%On the three days before the TeV gamma-ray flare, the optical EVPA temporarily became nearly aligned with the PA of the jet, before it suddenly rotated on the day before the TeV gamma-ray flare. 
During the three days preceding the TeV gamma-ray flare, the optical EVPA became (temporarily) nearly aligned with the PA of the jet, but on the day before the TeV gamma-ray flare it suddenly rotated back to a direction consistent with its direction prior to this quasi-alignment. 
Such abrupt changes in optical polarization associated with flares are found in numerical simulations for blazars and gamma-ray bursts \citep{Zhang14, Deng16}, 
%It can potentially be interpreted as 
and can potentially be interpreted as resulting from the helical motion of an emitting component in a toroidal/helical magnetic field before that component reaches the shocked region \citep[e.g.,][]{Marscher08}. 
However, since the fractional polarization was relatively low during this period, it is also possible that the observed EVPA change was a random fluctuation due to a turbulent magnetic field. 
%This may be explained by a passing perturbation and the consequently less organized magnetic field structure in the jet. % caused by $K16$. 

%Flare is consistent with magnetic reconnection. 
%Destructuration of the jet: a magnetic reconnection event can break the MHD jet structure 
A superluminal radio knot $K16$ was observed through a series of VLBA exposures on BL~Lacertae at 43 GHz. Extrapolation of the knot position implies that the VHE gamma-ray flare happened as the knot $K16$ crossed the quasi-stationary radio core. 
This suggests a possible association between the fast VHE gamma-ray flare and the emergence of the superluminal radio knot for the source, similar to that reported by \citet{Arlen13}. 
%This is the second detection of a fast TeV gamma-ray flare and the appearance of a superluminal radio knot coincidental in time from BL~Lacertae, making it less likely to be a random coincidence. 

\subsection{Interpretations of the TeV gamma-ray and the radio results}

In the model proposed by \citet{Marscher14}, the radio core is a Mach disk at the apex of a conical shock downstream in the jet, with a transverse orientation with respect to the jet axis. 
%When turbulent cells of plasma pass through the conical shock, electrons can be accelerated to relativistic speed. %and produce a gamma-ray flare through inverse-Compton scattering. 
%A fast gamma-ray flare can happen via inverse-Compton scattering as the relativistic plasma passes the Mach disk at the end of the conical shock. 
%The slow but highly compressed plasma in the Mach disk provides a highly variable local source of seed photons for inverse Compton scattering by electrons in the faster plasma that passes across the conical shock. If a region of especially high density of relativistic electrons passes through the core, it can cause a sharp flare at gamma-ray energies and 
%After the plasma flows through the Mach disk, a conical rarefaction can cause the flow to expand and accelerate and appear as a superluminal radio knot. 
%
When turbulent cells of plasma pass through the conical shock, relativistic 
electrons can be accelerated to higher energies in those cells where the magnetic 
field orientation relative to the shock normal is favorable. A fast gamma-ray flare 
can happen via inverse-Compton scattering as the relativistic plasma approaches the
Mach disk at the end of the conical shock, which provides a dense source of 
synchrotron and SSC seed photons. After the energized plasma passes the Mach disk, a 
conical rarefaction causes the flow to expand and accelerate, with the bright plasma 
appearing as a superluminal radio knot. 

%This model is able to produce sharp flares, as well as variations in polarization angle and fraction, consistent with the observations. 
%Variations in the polarization including a drop in the polarization fraction (due to the new linear polarization of the passing plasma cancelling that of the stationary core) and/or a swing in the polarization angle (as the passing plasma becomes brighter and dominant over the stationary core) are predicted for gamma-ray flares in similar models 
In some numerical simulations, the polarization fraction drops as the magnetic field direction changes, while the EVPA can rotate owing to random fluctuation of the field or the emergence of a new field component \citep[e.g.,][]{Marscher14, Zhang14}. 
This is consistent with the variation observed in the $R$-band polarization shortly before the VHE gamma-ray flare (see Figure~\ref{fig:mLC1}), as well as the VLBA images at 15.4~GHz (see Figure~\ref{fig:VLBA15}). The changing superposition of the magnetic fields as the moving knot ($K16$) passes the quasi-stationary knots ($A0$, $A1$, and $A2$) may also explain the change in the positions of $A0$, $A1$, and $A2$ between epochs (see Figure~\ref{fig:VLBA}). 
%Although we note that in this model the flares are caused by continuous noise processes of turbulent instead of singular events such as explosive injection of relativistic particles at the base of the jet, in contrast with the model we invoked above \citep{Katarzynski03} to explain the sharp fall of the flux. 
%Such an association was supported by the optical and radio polarization evolution around the time of the flare. 

%
An alternative hypothesis that can explain both the VHE gamma-ray flare and the superluminal radio knot of BL~Lacertae %the break out of a stationary knot at a recollimation shock 
is the breakout of a recollimation-shock zone %(observed as a stationary knot before the breakout) 
\citep{Hervet16}. %, is also possible. 
In this model, one or more recollimation shocks, of similar nature to those in \citet{Marscher14} \citep[see also][]{Mizuno15, Fromm16}, can form upstream in the jet where the magnetic energy density is high and appear as stationary radio knots; further downstream in the jet, particle kinetic energy becomes dominant, the magnetic field becomes unstable, and a stationary knot can be carried away by the underlying relativistic flow and become a superluminal knot. 
In the case of a compact region with large kinetic energy passing the recollimation-shock zone, %that results in a multi-component flare with one component behaving like a constant baseline on timescales of hours. 
a multi-component flare could be observed, with one component that varies slowly (i.e., on timescales of hours), thereby giving the appearance of a quasi-constant baseline in an intra-night light curve, as a result of the following sequence of events. 
First, in this scenario, an increase in the non-thermal emission of the shock region is expected, which leads to a flux increase on the timescale corresponding to the size of the entire shock region (as the baseline component). 
As the kinetic power of the jet increases at the shock zone, %the relative strength of the magnetic field decreases until the magnetic field structure can no longer be supported, at which point a magnetic reconnection event occurs, leading to the observed fast flare. 
the magnetic field structure is subject to strong tearing instabilities, at which point a magnetic-reconnection event occurs, leading to the observed fast flare. 
Finally, the shock zone is dragged away by the flow and enters an adiabatic expansion and cooling phase, leading to a decrease in flux and a return to the low state of the source. 
In the case of the 2016 flare of BL~Lacertae, there is no evidence for any disruption or breakout of a stationary knot, although it is possible that the recollimation zone reformed quickly between VLBA epochs and was therefore not sampled by the observations. 
Therefore, future observations of flares from gamma-ray blazars, with adequate coverage after the flux decreases, can potentially reduce the ambiguity in the interpretation. 

%The current data do not allow a unique interpretation of the association between them. 

%The observed fast flare can also be explained by magnetic reconnection \citep{Giannios09}. In this model, material with high magnetization can occasionally advect into small regions within the bulk of a magnetically dominated jet, and dissipate energy through magnetic reconnection, forming a ``jet-in-a-jet'' that is relativistic in the jet frame. Such a reconnection region is observed in the lab frame at a much higher Lorentz factor compared to the bulk of the jet, and is able to produce a fast TeV gamma-ray flare through inverse-Compton process. Note that an simultaneous X-ray flare is naturally predicted in this model. 

%...(more after VLBA data are delivered)
%The emergence of a new component observed ??

%Where was the flare produced, we don't know. 
%Radio knows where. Maybe TeV is related to radio: the same blob manifests activities observed in both band. 

%Polarization in radio and optical support this relation. 

%
\acknowledgments
VERITAS is supported by grants from the U.S. Department of Energy Office of Science, the U.S. National Science Foundation and the Smithsonian Institution, and by NSERC in Canada. We acknowledge the excellent work of the technical support staff at the Fred Lawrence Whipple Observatory and at the collaborating institutions in the construction and operation of the instrument. 

The VERITAS Collaboration is grateful to Trevor Weekes for his seminal contributions and leadership in the field of VHE gamma-ray astrophysics, which made this study possible.

%The research at Boston University was supported in part by NASA \textit{Fermi} Guest Investigator Program grant NNX14AQ58G.
The research at Boston University was supported in part by NASA \textit{Fermi} Guest Investigator Program grant 80NSSC17K0694.
The VLBA is an instrument of the Long Baseline Observatory (LBO). The LBO is a facility of the National Science Foundation operated under cooperative agreement by Associated Universities, Inc. 

This research has made use of data from the MOJAVE database that is maintained by the MOJAVE team \citep{Lister09} and supported by NASA-{\it Fermi} grant NNX15AU76G. This work made use of the Swinburne University of Technology software correlator \citep{Deller11}, developed as part of the Australian Major National Research Facilities Programme and operated under licence. YYK and ABP are partly supported by the Russian Foundation for Basic Research (project 17-02-00197), the government of the Russian Federation (agreement 05.Y09.21.0018), and the Alexander von Humboldt Foundation. T.S. was funded by the Academy of Finland projects 274477 and 284495.

This research has made use of data from the OVRO 40-m monitoring program \citep{Richards11} which is supported in part by NASA grants NNX08AW31G, NNX11A043G, and NNX14AQ89G and NSF grants AST-0808050 and AST-1109911.

The monitoring of BL~Lacertae and other blazars at the Steward Observatory is supported through NASA \textit{Fermi} Guest Investigator grant NNX15AU81G.

IA acknowledges support by a Ram\'on y Cajal grant of the Ministerio de Econom\'ia y Competitividad (MINECO) of Spain. Acquisition and reduction of the MAPCAT data was supported in part by MINECO through grants AYA2010-14844, AYA2013-40825-P, and AYA2016-80889-P, and by the Regional Government of Andaluc\'ia through grant P09-FQM-4784. The MAPCAT observations were carried out at the German-Spanish Calar Alto Observatory, which is jointly operated by the Max-Plank-Institut f\"ur Astronomie and the Instituto de Astrof\'isica de Andaluc\'ia-CSIC. 

The St. Petersburg University team acknowledges support from Russian Science Foundation grant 17-12-01029.

\vspace{5mm}
\facilities{VERITAS, \textit{Fermi}(LAT), \textit{Swift}(XRT), SO:Kuiper, Bok, CrAO:1.25m, CAO:2.2m, Perkins, LX-200, FLWO:1.2m, VLBA,  Mets\"ahovi Radio Observatory, OVRO:40m}


\software{%astropy \citep{Astropy13},  
          Emcee \citep{Foreman-Mackey13}, 
          NumPy \citep{numpy11},
          Matplotlib \citep{Hunter07},
          SciPy \citep{SciPy},
          Seaborn \citep{seaborn14}
          }

\bibliography{QiBibAll}

\begin{thebibliography}{}
\expandafter\ifx\csname natexlab\endcsname\relax\def\natexlab#1{#1}\fi

\bibitem[{{Abdo} {et~al.}(2010){Abdo}, {Ackermann}, {Ajello}, {Axelsson},
  {Baldini}, {Ballet}, {Barbiellini}, {Bastieri}, {Baughman}, {Bechtol}, \&
  et~al.}]{Abdo10PolRot}
{Abdo}, A.~A., {Ackermann}, M., {Ajello}, M., {et~al.} 2010, \nat, 463, 919

\bibitem[{{Abdo} {et~al.}(2011){Abdo}, {Ackermann}, {Ajello}, {Antolini},
  {Baldini}, {Ballet}, {Barbiellini}, {Bastieri}, {Bechtol}, {Bellazzini},
  {Berenji}, {Blandford}, {Bonamente}, {Borgland}, {Bregeon}, {Brez},
  {Brigida}, {Bruel}, {Buehler}, {Buson}, {Caliandro}, {Cameron}, {Cannon},
  {Caraveo}, {Carrigan}, {Casandjian}, {Cecchi}, {{\c C}elik}, {Charles},
  {Chekhtman}, {Cheung}, {Chiang}, {Ciprini}, {Claus}, {Cohen-Tanugi},
  {Conrad}, {Costamante}, {Cutini}, {Dermer}, {de Palma}, {Donato}, {Silva},
  {Drell}, {Dubois}, {Escande}, {Favuzzi}, {Fegan}, {Finke}, {Focke}, {Fortin},
  {Frailis}, {Fukazawa}, {Funk}, {Fusco}, {Gargano}, {Gasparrini}, {Gehrels},
  {Germani}, {Giglietto}, {Giordano}, {Giroletti}, {Glanzman}, {Godfrey},
  {Grenier}, {Guiriec}, {Hadasch}, {Hayashida}, {Hays}, {Hughes}, {Itoh},
  {J{\'o}hannesson}, {Johnson}, {Johnson}, {Kamae}, {Katagiri}, {Kataoka},
  {Kn{\"o}dlseder}, {Kuss}, {Lande}, {Larsson}, {Latronico}, {Lee}, {Llena
  Garde}, {Longo}, {Loparco}, {Lott}, {Lovellette}, {Lubrano}, {Makeev},
  {Mazziotta}, {McEnery}, {Mehault}, {Michelson}, {Mizuno}, {Monte}, {Monzani},
  {Morselli}, {Moskalenko}, {Murgia}, {Nakamori}, {Naumann-Godo}, {Nishino},
  {Nolan}, {Norris}, {Nuss}, {Ohsugi}, {Okumura}, {Omodei}, {Orlando}, {Ormes},
  {Ozaki}, {Paneque}, {Panetta}, {Parent}, {Pelassa}, {Pepe}, {Pesce-Rollins},
  {Piron}, {Porter}, {Rain{\`o}}, {Rando}, {Razzano}, {Reimer}, {Reimer},
  {Ritz}, {Roth}, {Sadrozinski}, {Sanchez}, {Sander}, {Schinzel}, {Sgr{\`o}},
  {Siskind}, {Smith}, {Sokolovsky}, {Spandre}, {Spinelli}, {Strickman},
  {Suson}, {Takahashi}, {Tanaka}, {Thayer}, {Thayer}, {Thompson}, {Tibaldo},
  {Torres}, {Tosti}, {Tramacere}, {Uehara}, {Usher}, {Vandenbroucke},
  {Vasileiou}, {Vilchez}, {Vitale}, {Waite}, {Wallace}, {Wang}, {Winer},
  {Wood}, {Yang}, {Ylinen}, {Ziegler}, {Berdyugin}, {Boettcher},
  {Carrami{\~n}ana}, {Carrasco}, {de la Fuente}, {Diltz}, {Hovatta},
  {Kadenius}, {Kovalev}, {L{\"a}hteenm{\"a}ki}, {Lindfors}, {Marscher},
  {Nilsson}, {Pereira}, {Reinthal}, {Roustazadeh}, {Savolainen},
  {Sillanp{\"a}{\"a}}, {Takalo}, \& {Tornikoski}}]{Abdo11BLLac}
---. 2011, \apj, 730, 101

\bibitem[{{Abeysekara} {et~al.}(2017){Abeysekara}, {Archambault}, {Archer},
  {Benbow}, {Bird}, {Buchovecky}, {Buckley}, {Bugaev}, {Cardenzana}, {Cerruti},
  \& et~al.}]{Abeysekara17}
{Abeysekara}, A.~U., {Archambault}, S., {Archer}, A., {et~al.} 2017, \apj, 834,
  2

\bibitem[{{Acero} {et~al.}(2015){Acero}, {Ackermann}, {Ajello}, {Albert},
  {Atwood}, {Axelsson}, {Baldini}, {Ballet}, {Barbiellini}, {Bastieri},
  {Belfiore}, {Bellazzini}, {Bissaldi}, {Blandford}, {Bloom}, {Bogart},
  {Bonino}, {Bottacini}, {Bregeon}, {Britto}, {Bruel}, {Buehler}, {Burnett},
  {Buson}, {Caliandro}, {Cameron}, {Caputo}, {Caragiulo}, {Caraveo},
  {Casandjian}, {Cavazzuti}, {Charles}, {Chaves}, {Chekhtman}, {Cheung},
  {Chiang}, {Chiaro}, {Ciprini}, {Claus}, {Cohen-Tanugi}, {Cominsky}, {Conrad},
  {Cutini}, {D'Ammando}, {de Angelis}, {DeKlotz}, {de Palma}, {Desiante},
  {Digel}, {Di Venere}, {Drell}, {Dubois}, {Dumora}, {Favuzzi}, {Fegan},
  {Ferrara}, {Finke}, {Franckowiak}, {Fukazawa}, {Funk}, {Fusco}, {Gargano},
  {Gasparrini}, {Giebels}, {Giglietto}, {Giommi}, {Giordano}, {Giroletti},
  {Glanzman}, {Godfrey}, {Grenier}, {Grondin}, {Grove}, {Guillemot}, {Guiriec},
  {Hadasch}, {Harding}, {Hays}, {Hewitt}, {Hill}, {Horan}, {Iafrate}, {Jogler},
  {J{\'o}hannesson}, {Johnson}, {Johnson}, {Johnson}, {Johnson}, {Kamae},
  {Kataoka}, {Katsuta}, {Kuss}, {La Mura}, {Landriu}, {Larsson}, {Latronico},
  {Lemoine-Goumard}, {Li}, {Li}, {Longo}, {Loparco}, {Lott}, {Lovellette},
  {Lubrano}, {Madejski}, {Massaro}, {Mayer}, {Mazziotta}, {McEnery},
  {Michelson}, {Mirabal}, {Mizuno}, {Moiseev}, {Mongelli}, {Monzani},
  {Morselli}, {Moskalenko}, {Murgia}, {Nuss}, {Ohno}, {Ohsugi}, {Omodei},
  {Orienti}, {Orlando}, {Ormes}, {Paneque}, {Panetta}, {Perkins},
  {Pesce-Rollins}, {Piron}, {Pivato}, {Porter}, {Racusin}, {Rando}, {Razzano},
  {Razzaque}, {Reimer}, {Reimer}, {Reposeur}, {Rochester}, {Romani},
  {Salvetti}, {S{\'a}nchez-Conde}, {Saz Parkinson}, {Schulz}, {Siskind},
  {Smith}, {Spada}, {Spandre}, {Spinelli}, {Stephens}, {Strong}, {Suson},
  {Takahashi}, {Takahashi}, {Tanaka}, {Thayer}, {Thayer}, {Thompson},
  {Tibaldo}, {Tibolla}, {Torres}, {Torresi}, {Tosti}, {Troja}, {Van Klaveren},
  {Vianello}, {Winer}, {Wood}, {Wood}, {Zimmer}, \& {Fermi-LAT
  Collaboration}}]{Acero15}
{Acero}, F., {Ackermann}, M., {Ajello}, M., {et~al.} 2015, \apjs, 218, 23

\bibitem[{{Ackermann} {et~al.}(2011){Ackermann}, {Ajello}, {Allafort},
  {Antolini}, {Atwood}, {Axelsson}, {Baldini}, {Ballet}, {Barbiellini},
  {Bastieri}, {Bechtol}, {Bellazzini}, {Berenji}, {Blandford}, {Bloom},
  {Bonamente}, {Borgland}, {Bottacini}, {Bouvier}, {Bregeon}, {Brigida},
  {Bruel}, {Buehler}, {Burnett}, {Buson}, {Caliandro}, {Cameron}, {Caraveo},
  {Casandjian}, {Cavazzuti}, {Cecchi}, {Charles}, {Cheung}, {Chiang},
  {Ciprini}, {Claus}, {Cohen-Tanugi}, {Conrad}, {Costamante}, {Cutini}, {de
  Angelis}, {de Palma}, {Dermer}, {Digel}, {Silva}, {Drell}, {Dubois},
  {Escande}, {Favuzzi}, {Fegan}, {Ferrara}, {Finke}, {Focke}, {Fortin},
  {Frailis}, {Fukazawa}, {Funk}, {Fusco}, {Gargano}, {Gasparrini}, {Gehrels},
  {Germani}, {Giebels}, {Giglietto}, {Giommi}, {Giordano}, {Giroletti},
  {Glanzman}, {Godfrey}, {Grenier}, {Grove}, {Guiriec}, {Gustafsson},
  {Hadasch}, {Hayashida}, {Hays}, {Healey}, {Horan}, {Hou}, {Hughes},
  {Iafrate}, {J{\'o}hannesson}, {Johnson}, {Johnson}, {Kamae}, {Katagiri},
  {Kataoka}, {Kn{\"o}dlseder}, {Kuss}, {Lande}, {Larsson}, {Latronico},
  {Longo}, {Loparco}, {Lott}, {Lovellette}, {Lubrano}, {Madejski}, {Mazziotta},
  {McConville}, {McEnery}, {Michelson}, {Mitthumsiri}, {Mizuno}, {Moiseev},
  {Monte}, {Monzani}, {Moretti}, {Morselli}, {Moskalenko}, {Murgia},
  {Nakamori}, {Naumann-Godo}, {Nolan}, {Norris}, {Nuss}, {Ohno}, {Ohsugi},
  {Okumura}, {Omodei}, {Orienti}, {Orlando}, {Ormes}, {Ozaki}, {Paneque},
  {Parent}, {Pesce-Rollins}, {Pierbattista}, {Piranomonte}, {Piron}, {Pivato},
  {Porter}, {Rain{\`o}}, {Rando}, {Razzano}, {Razzaque}, {Reimer}, {Reimer},
  {Ritz}, {Rochester}, {Romani}, {Roth}, {Sanchez}, {Sbarra}, {Scargle},
  {Schalk}, {Sgr{\`o}}, {Shaw}, {Siskind}, {Spandre}, {Spinelli}, {Strong},
  {Suson}, {Tajima}, {Takahashi}, {Takahashi}, {Tanaka}, {Thayer}, {Thayer},
  {Thompson}, {Tibaldo}, {Tinivella}, {Torres}, {Tosti}, {Troja}, {Uchiyama},
  {Vandenbroucke}, {Vasileiou}, {Vianello}, {Vitale}, {Waite}, {Wallace},
  {Wang}, {Winer}, {Wood}, {Wood}, \& {Zimmer}}]{Ackermann11}
{Ackermann}, M., {Ajello}, M., {Allafort}, A., {et~al.} 2011, \apj, 743, 171

\bibitem[{{Aharonian} {et~al.}(2007){Aharonian}, {Akhperjanian}, {Bazer-Bachi},
  {Behera}, {Beilicke}, {Benbow}, {Berge}, {Bernl{\"o}hr}, {Boisson}, {Bolz},
  {Borrel}, {Boutelier}, {Braun}, {Brion}, {Brown}, {B{\"u}hler},
  {B{\"u}sching}, {Bulik}, {Carrigan}, {Chadwick}, {Clapson}, {Chounet},
  {Coignet}, {Cornils}, {Costamante}, {Degrange}, {Dickinson},
  {Djannati-Ata{\"i}}, {Domainko}, {Drury}, {Dubus}, {Dyks}, {Egberts},
  {Emmanoulopoulos}, {Espigat}, {Farnier}, {Feinstein}, {Fiasson},
  {F{\"o}rster}, {Fontaine}, {Funk}, {Funk}, {F{\"u}{\ss}ling}, {Gallant},
  {Giebels}, {Glicenstein}, {Gl{\"u}ck}, {Goret}, {Hadjichristidis}, {Hauser},
  {Hauser}, {Heinzelmann}, {Henri}, {Hermann}, {Hinton}, {Hoffmann}, {Hofmann},
  {Holleran}, {Hoppe}, {Horns}, {Jacholkowska}, {de Jager}, {Kendziorra},
  {Kerschhaggl}, {Kh{\'e}lifi}, {Komin}, {Kosack}, {Lamanna}, {Latham}, {Le
  Gallou}, {Lemi{\`e}re}, {Lemoine-Goumard}, {Lenain}, {Lohse}, {Martin},
  {Martineau-Huynh}, {Marcowith}, {Masterson}, {Maurin}, {McComb}, {Moderski},
  {Moulin}, {de Naurois}, {Nedbal}, {Nolan}, {Olive}, {Orford}, {Osborne},
  {Ostrowski}, {Panter}, {Pedaletti}, {Pelletier}, {Petrucci}, {Pita},
  {P{\"u}hlhofer}, {Punch}, {Ranchon}, {Raubenheimer}, {Raue}, {Rayner},
  {Renaud}, {Ripken}, {Rob}, {Rolland}, {Rosier-Lees}, {Rowell}, {Rudak},
  {Ruppel}, {Sahakian}, {Santangelo}, {Saug{\'e}}, {Schlenker}, {Schlickeiser},
  {Schr{\"o}der}, {Schwanke}, {Schwarzburg}, {Schwemmer}, {Shalchi}, {Sol},
  {Spangler}, {Stawarz}, {Steenkamp}, {Stegmann}, {Superina}, {Tam},
  {Tavernet}, {Terrier}, {van Eldik}, {Vasileiadis}, {Venter}, {Vialle},
  {Vincent}, {Vivier}, {V{\"o}lk}, {Volpe}, {Wagner}, {Ward}, \&
  {Zdziarski}}]{Aharonian07}
{Aharonian}, F., {Akhperjanian}, A.~G., {Bazer-Bachi}, A.~R., {et~al.} 2007,
  \apjl, 664, L71

\bibitem[{{Albert} {et~al.}(2007{\natexlab{a}}){Albert}, {Aliu}, {Anderhub},
  {Antoranz}, {Armada}, {Baixeras}, {Barrio}, {Bartko}, {Bastieri}, {Becker},
  {Bednarek}, {Berger}, {Bigongiari}, {Biland}, {Bock}, {Bordas},
  {Bosch-Ramon}, {Bretz}, {Britvitch}, {Camara}, {Carmona}, {Chilingarian},
  {Coarasa}, {Commichau}, {Contreras}, {Cortina}, {Costado}, {Curtef},
  {Danielyan}, {Dazzi}, {De Angelis}, {Delgado}, {de los Reyes}, {De Lotto},
  {Domingo-Santamar{\'{\i}}a}, {Dorner}, {Doro}, {Errando}, {Fagiolini},
  {Ferenc}, {Fern{\'a}ndez}, {Firpo}, {Flix}, {Fonseca}, {Font}, {Fuchs},
  {Galante}, {Garc{\'{\i}}a-L{\'o}pez}, {Garczarczyk}, {Gaug}, {Giller},
  {Goebel}, {Hakobyan}, {Hayashida}, {Hengstebeck}, {Herrero}, {H{\"o}hne},
  {Hose}, {Hsu}, {Jacon}, {Jogler}, {Kosyra}, {Kranich}, {Kritzer}, {Laille},
  {Lindfors}, {Lombardi}, {Longo}, {L{\'o}pez}, {L{\'o}pez}, {Lorenz},
  {Majumdar}, {Maneva}, {Mannheim}, {Mansutti}, {Mariotti}, {Mart{\'{\i}}nez},
  {Mazin}, {Merck}, {Meucci}, {Meyer}, {Miranda}, {Mirzoyan}, {Mizobuchi},
  {Moralejo}, {Nilsson}, {Ninkovic}, {O{\~n}a-Wilhelmi}, {Otte}, {Oya},
  {Paneque}, {Panniello}, {Paoletti}, {Paredes}, {Pasanen}, {Pascoli}, {Pauss},
  {Pegna}, {Persic}, {Peruzzo}, {Piccioli}, {Poller}, {Prandini}, {Puchades},
  {Raymers}, {Rhode}, {Rib{\'o}}, {Rico}, {Rissi}, {Robert}, {R{\"u}gamer},
  {Saggion}, {S{\'a}nchez}, {Sartori}, {Scalzotto}, {Scapin}, {Schmitt},
  {Schweizer}, {Shayduk}, {Shinozaki}, {Shore}, {Sidro}, {Sillanp{\"a}{\"a}},
  {Sobczynska}, {Stamerra}, {Stark}, {Takalo}, {Temnikov}, {Tescaro},
  {Teshima}, {Tonello}, {Torres}, {Turini}, {Vankov}, {Vitale}, {Wagner},
  {Wibig}, {Wittek}, {Zandanel}, {Zanin}, \& {Zapatero}}]{Albert07BLLac}
{Albert}, J., {Aliu}, E., {Anderhub}, H., {et~al.} 2007{\natexlab{a}}, \apjl,
  666, L17

\bibitem[{{Albert} {et~al.}(2007{\natexlab{b}}){Albert}, {Aliu}, {Anderhub},
  {Antoranz}, {Armada}, {Baixeras}, {Barrio}, {Bartko}, {Bastieri}, {Becker},
  {Bednarek}, {Berger}, {Bigongiari}, {Biland}, {Bock}, {Bordas},
  {Bosch-Ramon}, {Bretz}, {Britvitch}, {Camara}, {Carmona}, {Chilingarian},
  {Coarasa}, {Commichau}, {Contreras}, {Cortina}, {Costado}, {Curtef},
  {Danielyan}, {Dazzi}, {De Angelis}, {Delgado}, {de los Reyes}, {De Lotto},
  {Domingo-Santamar{\'{\i}}a}, {Dorner}, {Doro}, {Errando}, {Fagiolini},
  {Ferenc}, {Fern{\'a}ndez}, {Firpo}, {Flix}, {Fonseca}, {Font}, {Fuchs},
  {Galante}, {Garc{\'{\i}}a-L{\'o}pez}, {Garczarczyk}, {Gaug}, {Giller},
  {Goebel}, {Hakobyan}, {Hayashida}, {Hengstebeck}, {Herrero}, {H{\"o}hne},
  {Hose}, {Hrupec}, {Hsu}, {Jacon}, {Jogler}, {Kosyra}, {Kranich}, {Kritzer},
  {Laille}, {Lindfors}, {Lombardi}, {Longo}, {L{\'o}pez}, {L{\'o}pez},
  {Lorenz}, {Majumdar}, {Maneva}, {Mannheim}, {Mansutti}, {Mariotti},
  {Mart{\'{\i}}nez}, {Mazin}, {Merck}, {Meucci}, {Meyer}, {Miranda},
  {Mirzoyan}, {Mizobuchi}, {Moralejo}, {Nieto}, {Nilsson}, {Ninkovic},
  {O{\~n}a-Wilhelmi}, {Otte}, {Oya}, {Paneque}, {Panniello}, {Paoletti},
  {Paredes}, {Pasanen}, {Pascoli}, {Pauss}, {Pegna}, {Persic}, {Peruzzo},
  {Piccioli}, {Prandini}, {Puchades}, {Raymers}, {Rhode}, {Rib{\'o}}, {Rico},
  {Rissi}, {Robert}, {R{\"u}gamer}, {Saggion}, {Saito}, {S{\'a}nchez},
  {Sartori}, {Scalzotto}, {Scapin}, {Schmitt}, {Schweizer}, {Shayduk},
  {Shinozaki}, {Shore}, {Sidro}, {Sillanp{\"a}{\"a}}, {Sobczynska}, {Stamerra},
  {Stark}, {Takalo}, {Tavecchio}, {Temnikov}, {Tescaro}, {Teshima}, {Torres},
  {Turini}, {Vankov}, {Vitale}, {Wagner}, {Wibig}, {Wittek}, {Zandanel},
  {Zanin}, \& {Zapatero}}]{Albert07}
---. 2007{\natexlab{b}}, \apj, 669, 862

\bibitem[{{Aleksi{\'c}} {et~al.}(2011){Aleksi{\'c}}, {Antonelli}, {Antoranz},
  {Backes}, {Barrio}, {Bastieri}, {Becerra Gonz{\'a}lez}, {Bednarek},
  {Berdyugin}, {Berger}, {Bernardini}, {Biland}, {Blanch}, {Bock}, {Boller},
  {Bonnoli}, {Borla Tridon}, {Braun}, {Bretz}, {Ca{\~n}ellas}, {Carmona},
  {Carosi}, {Colin}, {Colombo}, {Contreras}, {Cortina}, {Cossio}, {Covino},
  {Dazzi}, {De Angelis}, {De Cea del Pozo}, {De Lotto}, {Delgado Mendez},
  {Diago Ortega}, {Doert}, {Dom{\'{\i}}nguez}, {Dominis Prester}, {Dorner},
  {Doro}, {Elsaesser}, {Ferenc}, {Fonseca}, {Font}, {Fruck}, {Garc{\'{\i}}a
  L{\'o}pez}, {Garczarczyk}, {Garrido}, {Giavitto}, {Godinovi{\'c}}, {Hadasch},
  {H{\"a}fner}, {Herrero}, {Hildebrand}, {H{\"o}hne-M{\"o}nch}, {Hose},
  {Hrupec}, {Huber}, {Jogler}, {Klepser}, {Kr{\"a}henb{\"u}hl}, {Krause}, {La
  Barbera}, {Lelas}, {Leonardo}, {Lindfors}, {Lombardi}, {L{\'o}pez}, {Lorenz},
  {Makariev}, {Maneva}, {Mankuzhiyil}, {Mannheim}, {Maraschi}, {Mariotti},
  {Mart{\'{\i}}nez}, {Mazin}, {Meucci}, {Miranda}, {Mirzoyan}, {Miyamoto},
  {Mold{\'o}n}, {Moralejo}, {Nieto}, {Nilsson}, {Orito}, {Oya}, {Paneque},
  {Paoletti}, {Pardo}, {Paredes}, {Partini}, {Pasanen}, {Pauss},
  {Perez-Torres}, {Persic}, {Peruzzo}, {Pilia}, {Pochon}, {Prada}, {Prada
  Moroni}, {Prandini}, {Puljak}, {Reichardt}, {Reinthal}, {Rhode}, {Rib{\'o}},
  {Rico}, {R{\"u}gamer}, {Saggion}, {Saito}, {Saito}, {Salvati}, {Satalecka},
  {Scalzotto}, {Scapin}, {Schultz}, {Schweizer}, {Shayduk}, {Shore},
  {Sillanp{\"a}{\"a}}, {Sitarek}, {Sobczynska}, {Spanier}, {Spiro}, {Stamerra},
  {Steinke}, {Storz}, {Strah}, {Suri{\'c}}, {Takalo}, {Tavecchio}, {Temnikov},
  {Terzi{\'c}}, {Tescaro}, {Teshima}, {Thom}, {Tibolla}, {Torres}, {Treves},
  {Vankov}, {Vogler}, {Wagner}, {Weitzel}, {Zabalza}, {Zandanel}, {Zanin},
  {MAGIC Collaboration}, {Tanaka}, {Wood}, \& {Buson}}]{Aleksic11}
{Aleksi{\'c}}, J., {Antonelli}, L.~A., {Antoranz}, P., {et~al.} 2011, \apjl,
  730, L8

\bibitem[{{Alexander}(2013)}]{Alexander13}
{Alexander}, T. 2013, ArXiv e-prints, arXiv:1302.1508

\bibitem[{{Algaba} {et~al.}(2011){Algaba}, {Gabuzda}, \& {Smith}}]{Algaba11}
{Algaba}, J.~C., {Gabuzda}, D.~C., \& {Smith}, P.~S. 2011, \mnras, 411, 85

\bibitem[{{Allen} {et~al.}(1982){Allen}, {Ward}, \& {Hyland}}]{Allen82}
{Allen}, D.~A., {Ward}, M.~J., \& {Hyland}, A.~R. 1982, \mnras, 199, 969

\bibitem[{{Archambault} {et~al.}(2014){Archambault}, {Aune}, {Behera},
  {Beilicke}, {Benbow}, {Berger}, {Bird}, {Biteau}, {Bugaev}, {Byrum},
  {Cardenzana}, {Cerruti}, {Chen}, {Ciupik}, {Connolly}, {Cui}, {Dumm},
  {Errando}, {Falcone}, {Federici}, {Feng}, {Finley}, {Fleischhack}, {Fortson},
  {Furniss}, {Galante}, {Gillanders}, {Griffin}, {Griffiths}, {Grube}, {Gyuk},
  {Hanna}, {Holder}, {Hughes}, {Humensky}, {Johnson}, {Kaaret}, {Kertzman},
  {Khassen}, {Kieda}, {Krawczynski}, {Krennrich}, {Kumar}, {Lang}, {Madhavan},
  {Maier}, {McCann}, {Meagher}, {Moriarty}, {Mukherjee}, {Nieto},
  {O'Faol{\'a}in de Bhr{\'o}ithe}, {Ong}, {Otte}, {Park}, {Pohl}, {Popkow},
  {Prokoph}, {Quinn}, {Ragan}, {Rajotte}, {Reyes}, {Reynolds}, {Richards},
  {Roache}, {Sembroski}, {Shahinyan}, {Staszak}, {Telezhinsky}, {Tucci},
  {Tyler}, {Varlotta}, {Vassiliev}, {Vincent}, {Wakely}, {Weinstein},
  {Welsing}, {Wilhelm}, {Williams}, {VERITAS Collaboration}, {Ackermann},
  {Ajello}, {Albert}, {Baldini}, {Bastieri}, {Bellazzini}, {Bissaldi},
  {Bregeon}, {Buehler}, {Buson}, {Caliandro}, {Cameron}, {Caraveo},
  {Cavazzuti}, {Charles}, {Chiang}, {Ciprini}, {Claus}, {Cutini}, {D'Ammando},
  {de Angelis}, {de Palma}, {Dermer}, {Digel}, {Di Venere}, {Drell}, {Favuzzi},
  {Franckowiak}, {Fusco}, {Gargano}, {Gasparrini}, {Giglietto}, {Giordano},
  {Giroletti}, {Grenier}, {Guiriec}, {Jogler}, {Kuss}, {Larsson}, {Latronico},
  {Longo}, {Loparco}, {Lubrano}, {Madejski}, {Mayer}, {Mazziotta}, {Michelson},
  {Mizuno}, {Monzani}, {Morselli}, {Murgia}, {Nuss}, {Ohsugi}, {Ormes},
  {Paneque}, {Perkins}, {Piron}, {Pivato}, {Rain{\`o}}, {Razzano}, {Reimer},
  {Reimer}, {Ritz}, {Schaal}, {Sgr{\`o}}, {Siskind}, {Spinelli}, {Takahashi},
  {Tibaldo}, {Tinivella}, {Troja}, {Vianello}, {Werner}, {Wood}, \& {Fermi LAT
  Collaboration}}]{Archambault14}
{Archambault}, S., {Aune}, T., {Behera}, B., {et~al.} 2014, \apjl, 785, L16

\bibitem[{{Arlen} {et~al.}(2013){Arlen}, {Aune}, {Beilicke}, {Benbow},
  {Bouvier}, {Buckley}, {Bugaev}, {Cesarini}, {Ciupik}, {Connolly}, {Cui},
  {Dickherber}, {Dumm}, {Errando}, {Falcone}, {Federici}, {Feng}, {Finley},
  {Finnegan}, {Fortson}, {Furniss}, {Galante}, {Gall}, {Griffin}, {Grube},
  {Gyuk}, {Hanna}, {Holder}, {Humensky}, {Kaaret}, {Karlsson}, {Kertzman},
  {Khassen}, {Kieda}, {Krawczynski}, {Krennrich}, {Maier}, {Moriarty},
  {Mukherjee}, {Nelson}, {O'Faol{\'a}in de Bhr{\'o}ithe}, {Ong}, {Orr}, {Park},
  {Perkins}, {Pichel}, {Pohl}, {Prokoph}, {Quinn}, {Ragan}, {Reyes},
  {Reynolds}, {Roache}, {Saxon}, {Schroedter}, {Sembroski}, {Staszak},
  {Telezhinsky}, {Te{\v s}i{\'c}}, {Theiling}, {Tsurusaki}, {Varlotta},
  {Vincent}, {Wakely}, {Weekes}, {Weinstein}, {Welsing}, {Williams}, {Zitzer},
  {VERITAS Collaboration}, {Jorstad}, {MacDonald}, {Marscher}, {Smith},
  {Walker}, {Hovatta}, {Richards}, {Max-Moerbeck}, {Readhead}, {Lister},
  {Kovalev}, {Pushkarev}, {Gurwell}, {L{\"a}hteenm{\"a}ki}, {Nieppola},
  {Tornikoski}, \& {J{\"a}rvel{\"a}}}]{Arlen13}
{Arlen}, T., {Aune}, T., {Beilicke}, M., {et~al.} 2013, \apj, 762, 92

\bibitem[{{Atwood} {et~al.}(2013){Atwood}, {Albert}, {Baldini}, {Tinivella},
  {Bregeon}, {Pesce-Rollins}, {Sgr{\`o}}, {Bruel}, {Charles}, {Drlica-Wagner},
  {Franckowiak}, {Jogler}, {Rochester}, {Usher}, {Wood}, {Cohen-Tanugi}, \&
  {S.~Zimmer for the Fermi-LAT Collaboration}}]{Atwood13}
{Atwood}, W., {Albert}, A., {Baldini}, L., {et~al.} 2013, ArXiv e-prints,
  arXiv:1303.3514

\bibitem[{{Atwood} {et~al.}(2009){Atwood}, {Abdo}, {Ackermann}, {Althouse},
  {Anderson}, {Axelsson}, {Baldini}, {Ballet}, {Band}, {Barbiellini}, \&
  et~al.}]{Atwood09}
{Atwood}, W.~B., {Abdo}, A.~A., {Ackermann}, M., {et~al.} 2009, \apj, 697, 1071

\bibitem[{{Baars} {et~al.}(1977){Baars}, {Genzel}, {Pauliny-Toth}, \&
  {Witzel}}]{Baars77}
{Baars}, J.~W.~M., {Genzel}, R., {Pauliny-Toth}, I.~I.~K., \& {Witzel}, A.
  1977, \aap, 61, 99

\bibitem[{{Bach} {et~al.}(2006){Bach}, {Villata}, {Raiteri}, {Agudo}, {Aller},
  {Aller}, {Denn}, {G{\'o}mez}, {Jorstad}, {Marscher}, {Mutel}, \&
  {Ter{\"a}sranta}}]{Bach06}
{Bach}, U., {Villata}, M., {Raiteri}, C.~M., {et~al.} 2006, \aap, 456, 105

\bibitem[{{Blandford} \& {Rees}(1978)}]{BlandfordRees78}
{Blandford}, R.~D., \& {Rees}, M.~J. 1978, \physscr, 17, 265

\bibitem[{{B{\"o}ttcher} \& {Chiang}(2002)}]{Bottcher02}
{B{\"o}ttcher}, M., \& {Chiang}, J. 2002, \apj, 581, 127

\bibitem[{{Burrows} {et~al.}(2005){Burrows}, {Hill}, {Nousek}, {Kennea},
  {Wells}, {Osborne}, {Abbey}, {Beardmore}, {Mukerjee}, {Short}, {Chincarini},
  {Campana}, {Citterio}, {Moretti}, {Pagani}, {Tagliaferri}, {Giommi},
  {Capalbi}, {Tamburelli}, {Angelini}, {Cusumano}, {Br{\"a}uninger}, {Burkert},
  \& {Hartner}}]{Burrows05}
{Burrows}, D.~N., {Hill}, J.~E., {Nousek}, J.~A., {et~al.} 2005, \ssr, 120, 165

\bibitem[{{Cogan}(2008)}]{Cogan08}
{Cogan}, P. 2008, in International Cosmic Ray Conference, Vol.~3, International
  Cosmic Ray Conference, 1385--1388

\bibitem[{{Cohen} {et~al.}(2007){Cohen}, {Lister}, {Homan}, {Kadler},
  {Kellermann}, {Kovalev}, \& {Vermeulen}}]{Cohen07}
{Cohen}, M.~H., {Lister}, M.~L., {Homan}, D.~C., {et~al.} 2007, \apj, 658, 232

\bibitem[{{Cohen} {et~al.}(2014){Cohen}, {Meier}, {Arshakian}, {Homan},
  {Hovatta}, {Kovalev}, {Lister}, {Pushkarev}, {Richards}, \&
  {Savolainen}}]{Cohen14}
{Cohen}, M.~H., {Meier}, D.~L., {Arshakian}, T.~G., {et~al.} 2014, \apj, 787,
  151

\bibitem[{{Daniel}(2008)}]{Daniel08}
{Daniel}, M.~K. 2008, in International Cosmic Ray Conference, Vol.~3,
  International Cosmic Ray Conference, 1325--1328

\bibitem[{{D'arcangelo} {et~al.}(2009){D'arcangelo}, {Marscher}, {Jorstad},
  {Smith}, {Larionov}, {Hagen-Thorn}, {Williams}, {Gear}, {Clemens}, {Sarcia},
  {Grabau}, {Tollestrup}, {Buie}, {Taylor}, \& {Dunham}}]{D'arcangelo09}
{D'arcangelo}, F.~D., {Marscher}, A.~P., {Jorstad}, S.~G., {et~al.} 2009, \apj,
  697, 985

\bibitem[{{Deller} {et~al.}(2011){Deller}, {Brisken}, {Phillips}, {Morgan},
  {Alef}, {Cappallo}, {Middelberg}, {Romney}, {Rottmann}, {Tingay}, \&
  {Wayth}}]{Deller11}
{Deller}, A.~T., {Brisken}, W.~F., {Phillips}, C.~J., {et~al.} 2011, \pasp,
  123, 275

\bibitem[{{Deng} {et~al.}(2016){Deng}, {Zhang}, {Zhang}, \& {Li}}]{Deng16}
{Deng}, W., {Zhang}, H., {Zhang}, B., \& {Li}, H. 2016, \apjl, 821, L12

\bibitem[{{Dom{\'{\i}}nguez} {et~al.}(2011){Dom{\'{\i}}nguez}, {Primack},
  {Rosario}, {Prada}, {Gilmore}, {Faber}, {Koo}, {Somerville},
  {P{\'e}rez-Torres}, {P{\'e}rez-Gonz{\'a}lez}, {Huang}, {Davis},
  {Guhathakurta}, {Barmby}, {Conselice}, {Lozano}, {Newman}, \&
  {Cooper}}]{Dominguez11}
{Dom{\'{\i}}nguez}, A., {Primack}, J.~R., {Rosario}, D.~J., {et~al.} 2011,
  \mnras, 410, 2556

\bibitem[{{Dondi} \& {Ghisellini}(1995)}]{Dondi95}
{Dondi}, L., \& {Ghisellini}, G. 1995, \mnras, 273, 583

\bibitem[{{Fan} {et~al.}(1998){Fan}, {Xie}, {Pecontal}, {Pecontal}, \&
  {Copin}}]{Fan98}
{Fan}, J.~H., {Xie}, G.~Z., {Pecontal}, E., {Pecontal}, A., \& {Copin}, Y.
  1998, \apj, 507, 173

\bibitem[{{Foreman-Mackey} {et~al.}(2013){Foreman-Mackey}, {Hogg}, {Lang}, \&
  {Goodman}}]{Foreman-Mackey13}
{Foreman-Mackey}, D., {Hogg}, D.~W., {Lang}, D., \& {Goodman}, J. 2013, \pasp,
  125, 306

\bibitem[{{Fromm} {et~al.}(2016){Fromm}, {Perucho}, {Mimica}, \&
  {Ros}}]{Fromm16}
{Fromm}, C.~M., {Perucho}, M., {Mimica}, P., \& {Ros}, E. 2016, \aap, 588, A101

\bibitem[{{Gaidos} {et~al.}(1996){Gaidos}, {Akerlof}, {Biller}, {Boyle},
  {Breslin}, {Buckley}, {Carter-Lewis}, {Catanese}, {Cawley}, {Fegan},
  {Finley}, {Gordo}, {Hillas}, {Krennrich}, {Lamb}, {Lessard}, {McEnery},
  {Masterson}, {Mohanty}, {Moriarty}, {Quinn}, {Rodgers}, {Rose}, {Samuelson},
  {Schubnell}, {Sembroski}, {Srinivasan}, {Weekes}, {Wilson}, \&
  {Zweerink}}]{Gaidos96}
{Gaidos}, J.~A., {Akerlof}, C.~W., {Biller}, S., {et~al.} 1996, \nat, 383, 319

\bibitem[{{Gehrels} {et~al.}(2004){Gehrels}, {Chincarini}, {Giommi}, {Mason},
  {Nousek}, {Wells}, {White}, {Barthelmy}, {Burrows}, {Cominsky}, {Hurley},
  {Marshall}, {M{\'e}sz{\'a}ros}, {Roming}, {Angelini}, {Barbier}, {Belloni},
  {Campana}, {Caraveo}, {Chester}, {Citterio}, {Cline}, {Cropper}, {Cummings},
  {Dean}, {Feigelson}, {Fenimore}, {Frail}, {Fruchter}, {Garmire}, {Gendreau},
  {Ghisellini}, {Greiner}, {Hill}, {Hunsberger}, {Krimm}, {Kulkarni}, {Kumar},
  {Lebrun}, {Lloyd-Ronning}, {Markwardt}, {Mattson}, {Mushotzky}, {Norris},
  {Osborne}, {Paczynski}, {Palmer}, {Park}, {Parsons}, {Paul}, {Rees},
  {Reynolds}, {Rhoads}, {Sasseen}, {Schaefer}, {Short}, {Smale}, {Smith},
  {Stella}, {Tagliaferri}, {Takahashi}, {Tashiro}, {Townsley}, {Tueller},
  {Turner}, {Vietri}, {Voges}, {Ward}, {Willingale}, {Zerbi}, \&
  {Zhang}}]{Gehrels04}
{Gehrels}, N., {Chincarini}, G., {Giommi}, P., {et~al.} 2004, \apj, 611, 1005

\bibitem[{{Ghisellini} {et~al.}(1998){Ghisellini}, {Celotti}, {Fossati},
  {Maraschi}, \& {Comastri}}]{Ghisellini98}
{Ghisellini}, G., {Celotti}, A., {Fossati}, G., {Maraschi}, L., \& {Comastri},
  A. 1998, \mnras, 301, 451

\bibitem[{{Ghisellini} \& {Tavecchio}(2009)}]{Ghisellini09}
{Ghisellini}, G., \& {Tavecchio}, F. 2009, \mnras, 397, 985

\bibitem[{{Ghisellini} \& {Tavecchio}(2015)}]{Ghisellini15}
---. 2015, \mnras, 448, 1060

\bibitem[{{Ghisellini} {et~al.}(2005){Ghisellini}, {Tavecchio}, \&
  {Chiaberge}}]{Ghisellini05}
{Ghisellini}, G., {Tavecchio}, F., \& {Chiaberge}, M. 2005, \aap, 432, 401

\bibitem[{{Giannios} {et~al.}(2009){Giannios}, {Uzdensky}, \&
  {Begelman}}]{Giannios09}
{Giannios}, D., {Uzdensky}, D.~A., \& {Begelman}, M.~C. 2009, \mnras, 395, L29

\bibitem[{{G{\'o}mez} {et~al.}(2016){G{\'o}mez}, {Lobanov}, {Bruni}, {Kovalev},
  {Marscher}, {Jorstad}, {Mizuno}, {Bach}, {Sokolovsky}, {Anderson}, {Galindo},
  {Kardashev}, \& {Lisakov}}]{Gomez16}
{G{\'o}mez}, J.~L., {Lobanov}, A.~P., {Bruni}, G., {et~al.} 2016, \apj, 817, 96

\bibitem[{{Gupta} {et~al.}(2012){Gupta}, {Pandey}, {Singh}, {Rani}, {Pan},
  {Fan}, \& {Gupta}}]{Gupta12}
{Gupta}, S.~P., {Pandey}, U.~S., {Singh}, K., {et~al.} 2012, \na, 17, 8

\bibitem[{{Harra} {et~al.}(2016){Harra}, {Schrijver}, {Janvier}, {Toriumi},
  {Hudson}, {Matthews}, {Woods}, {Hara}, {Guedel}, {Kowalski}, {Osten},
  {Kusano}, \& {Lueftinger}}]{Harra16}
{Harra}, L.~K., {Schrijver}, C.~J., {Janvier}, M., {et~al.} 2016, \solphys,
  291, 1761

\bibitem[{{Hervet} {et~al.}(2015){Hervet}, {Boisson}, \& {Sol}}]{Hervet15}
{Hervet}, O., {Boisson}, C., \& {Sol}, H. 2015, \aap, 578, A69

\bibitem[{{Hervet} {et~al.}(2016){Hervet}, {Boisson}, \& {Sol}}]{Hervet16}
---. 2016, \aap, 592, A22

\bibitem[{{Holder}(2011)}]{Holder11}
{Holder}, J. 2011, International Cosmic Ray Conference, 12, 137

\bibitem[{{Hovatta} {et~al.}(2012){Hovatta}, {Lister}, {Aller}, {Aller},
  {Homan}, {Kovalev}, {Pushkarev}, \& {Savolainen}}]{Hovatta12}
{Hovatta}, T., {Lister}, M.~L., {Aller}, M.~F., {et~al.} 2012, \aj, 144, 105

\bibitem[{Hunter(2007)}]{Hunter07}
Hunter, J.~D. 2007, Computing in Science and Engineering, 9, 90

\bibitem[{Jones {et~al.}(2001)Jones, Oliphant, Peterson, {et~al.}}]{SciPy}
Jones, E., Oliphant, T., Peterson, P., {et~al.} 2001, {SciPy}: Open source
  scientific tools for {Python}, , , [Online; accessed 2017-02-10]

\bibitem[{{Jorstad} \& {Marscher}(2016)}]{Jorstad16}
{Jorstad}, S., \& {Marscher}, A. 2016, Galaxies, 4, 47

\bibitem[{{Jorstad} {et~al.}(2005){Jorstad}, {Marscher}, {Lister}, {Stirling},
  {Cawthorne}, {Gear}, {G{\'o}mez}, {Stevens}, {Smith}, {Forster}, \&
  {Robson}}]{Jorstad05}
{Jorstad}, S.~G., {Marscher}, A.~P., {Lister}, M.~L., {et~al.} 2005, \aj, 130,
  1418

\bibitem[{{Jorstad} {et~al.}(2007){Jorstad}, {Marscher}, {Stevens}, {Smith},
  {Forster}, {Gear}, {Cawthorne}, {Lister}, {Stirling}, {G{\'o}mez}, {Greaves},
  \& {Robson}}]{Jorstad07}
{Jorstad}, S.~G., {Marscher}, A.~P., {Stevens}, J.~A., {et~al.} 2007, \aj, 134,
  799

\bibitem[{{Jorstad} {et~al.}(2017){Jorstad}, {Marscher}, {Morozova},
  {Troitsky}, {Agudo}, {Casadio}, {Foord}, {G{\'o}mez}, {MacDonald}, {Molina},
  {L{\"a}hteenm{\"a}ki}, {Tammi}, \& {Tornikoski}}]{Jorstad17}
{Jorstad}, S.~G., {Marscher}, A.~P., {Morozova}, D.~A., {et~al.} 2017, \apj,
  846, 98

\bibitem[{{Kalberla} {et~al.}(2005){Kalberla}, {Burton}, {Hartmann}, {Arnal},
  {Bajaja}, {Morras}, \& {P{\"o}ppel}}]{Kalberla05}
{Kalberla}, P.~M.~W., {Burton}, W.~B., {Hartmann}, D., {et~al.} 2005, \aap,
  440, 775

\bibitem[{{Katarzy{\'n}ski} {et~al.}(2003){Katarzy{\'n}ski}, {Sol}, \&
  {Kus}}]{Katarzynski03}
{Katarzy{\'n}ski}, K., {Sol}, H., \& {Kus}, A. 2003, \aap, 410, 101

\bibitem[{{Kharb} {et~al.}(2008){Kharb}, {Lister}, \& {Shastri}}]{Kharb08}
{Kharb}, P., {Lister}, M.~L., \& {Shastri}, P. 2008, International Journal of
  Modern Physics D, 17, 1545

\bibitem[{{Kovalev} {et~al.}(2009){Kovalev}, {Aller}, {Aller}, {Homan},
  {Kadler}, {Kellermann}, {Kovalev}, {Lister}, {McCormick}, {Pushkarev}, {Ros},
  \& {Zensus}}]{Kovalev09}
{Kovalev}, Y.~Y., {Aller}, H.~D., {Aller}, M.~F., {et~al.} 2009, \apjl, 696,
  L17

\bibitem[{{Larson} {et~al.}(2011){Larson}, {Dunkley}, {Hinshaw}, {Komatsu},
  {Nolta}, {Bennett}, {Gold}, {Halpern}, {Hill}, {Jarosik}, {Kogut}, {Limon},
  {Meyer}, {Odegard}, {Page}, {Smith}, {Spergel}, {Tucker}, {Weiland},
  {Wollack}, \& {Wright}}]{Larson11}
{Larson}, D., {Dunkley}, J., {Hinshaw}, G., {et~al.} 2011, \apjs, 192, 16

\bibitem[{{Lister} {et~al.}(2009){Lister}, {Aller}, {Aller}, {Cohen}, {Homan},
  {Kadler}, {Kellermann}, {Kovalev}, {Ros}, {Savolainen}, {Zensus}, \&
  {Vermeulen}}]{Lister09}
{Lister}, M.~L., {Aller}, H.~D., {Aller}, M.~F., {et~al.} 2009, \aj, 137, 3718

\bibitem[{{Lister} {et~al.}(2013){Lister}, {Aller}, {Aller}, {Homan},
  {Kellermann}, {Kovalev}, {Pushkarev}, {Richards}, {Ros}, \&
  {Savolainen}}]{Lister13}
{Lister}, M.~L., {Aller}, M.~F., {Aller}, H.~D., {et~al.} 2013, \aj, 146, 120

\bibitem[{{Lister} {et~al.}(2016){Lister}, {Aller}, {Aller}, {Homan},
  {Kellermann}, {Kovalev}, {Pushkarev}, {Richards}, {Ros}, \&
  {Savolainen}}]{Lister16}
---. 2016, \aj, 152, 12

\bibitem[{{Madejski} \& {Sikora}(2016)}]{Madejski16}
{Madejski}, G.~., \& {Sikora}, M. 2016, \araa, 54, 725

\bibitem[{{Madejski} {et~al.}(1999){Madejski}, {Sikora}, {Jaffe},
  {B{\L}a{\.z}ejowski}, {Jahoda}, \& {Moderski}}]{Madejski99}
{Madejski}, G.~M., {Sikora}, M., {Jaffe}, T., {et~al.} 1999, \apj, 521, 145

\bibitem[{{Marscher}(2014)}]{Marscher14}
{Marscher}, A.~P. 2014, \apj, 780, 87

\bibitem[{{Marscher} {et~al.}(2008){Marscher}, {Jorstad}, {D'Arcangelo},
  {Smith}, {Williams}, {Larionov}, {Oh}, {Olmstead}, {Aller}, {Aller},
  {McHardy}, {L{\"a}hteenm{\"a}ki}, {Tornikoski}, {Valtaoja}, {Hagen-Thorn},
  {Kopatskaya}, {Gear}, {Tosti}, {Kurtanidze}, {Nikolashvili}, {Sigua},
  {Miller}, \& {Ryle}}]{Marscher08}
{Marscher}, A.~P., {Jorstad}, S.~G., {D'Arcangelo}, F.~D., {et~al.} 2008, \nat,
  452, 966

\bibitem[{{Max-Moerbeck} {et~al.}(2014){Max-Moerbeck}, {Hovatta}, {Richards},
  {King}, {Pearson}, {Readhead}, {Reeves}, {Shepherd}, {Stevenson},
  {Angelakis}, {Fuhrmann}, {Grainge}, {Pavlidou}, {Romani}, \&
  {Zensus}}]{Max-Moerbeck14}
{Max-Moerbeck}, W., {Hovatta}, T., {Richards}, J.~L., {et~al.} 2014, \mnras,
  445, 428

\bibitem[{{McLure} \& {Dunlop}(2002)}]{McLure02}
{McLure}, R.~J., \& {Dunlop}, J.~S. 2002, \mnras, 331, 795

\bibitem[{{Miller} \& {Hawley}(1977)}]{Miller77}
{Miller}, J.~S., \& {Hawley}, S.~A. 1977, \apjl, 212, L47

\bibitem[{{Mizuno} {et~al.}(2015){Mizuno}, {G{\'o}mez}, {Nishikawa}, {Meli},
  {Hardee}, \& {Rezzolla}}]{Mizuno15}
{Mizuno}, Y., {G{\'o}mez}, J.~L., {Nishikawa}, K.-I., {et~al.} 2015, \apj, 809,
  38

\bibitem[{{Moretti} {et~al.}(2005){Moretti}, {Campana}, {Mineo}, {Romano},
  {Abbey}, {Angelini}, {Beardmore}, {Burkert}, {Burrows}, {Capalbi},
  {Chincarini}, {Citterio}, {Cusumano}, {Freyberg}, {Giommi}, {Goad}, {Godet},
  {Hartner}, {Hill}, {Kennea}, {La Parola}, {Mangano}, {Morris}, {Nousek},
  {Osborne}, {Page}, {Pagani}, {Perri}, {Tagliaferri}, {Tamburelli}, \&
  {Wells}}]{Moretti05}
{Moretti}, A., {Campana}, S., {Mineo}, T., {et~al.} 2005, in \procspie, Vol.
  5898, UV, X-Ray, and Gamma-Ray Space Instrumentation for Astronomy XIV, ed.
  O.~H.~W. {Siegmund}, 360--368

\bibitem[{{Nalewajko} {et~al.}(2014){Nalewajko}, {Begelman}, \&
  {Sikora}}]{Nalewajko14}
{Nalewajko}, K., {Begelman}, M.~C., \& {Sikora}, M. 2014, \apj, 789, 161

\bibitem[{{Neshpor} {et~al.}(2001){Neshpor}, {Chalenko}, {Stepanian},
  {Kalekin}, {Jogolev}, {Fomin}, \& {Shitov}}]{Neshpor01}
{Neshpor}, Y.~I., {Chalenko}, N.~N., {Stepanian}, A.~A., {et~al.} 2001,
  Astronomy Reports, 45, 249

\bibitem[{{Petropoulou} {et~al.}(2016){Petropoulou}, {Giannios}, \&
  {Sironi}}]{Petropoulou16}
{Petropoulou}, M., {Giannios}, D., \& {Sironi}, L. 2016, \mnras, 462, 3325

\bibitem[{{Pollack} {et~al.}(2016){Pollack}, {Pauls}, \& {Wiita}}]{Pollack16}
{Pollack}, M., {Pauls}, D., \& {Wiita}, P.~J. 2016, \apj, 820, 12

\bibitem[{{Raiteri} {et~al.}(2013){Raiteri}, {Villata}, {D'Ammando},
  {Larionov}, {Gurwell}, {Mirzaqulov}, {Smith}, {Acosta-Pulido}, {Agudo},
  {Ar{\'e}valo}, {Bachev}, {Ben{\'{\i}}tez}, {Berdyugin}, {Blinov}, {Borman},
  {B{\"o}ttcher}, {Bozhilov}, {Carnerero}, {Carosati}, {Casadio}, {Chen},
  {Doroshenko}, {Efimov}, {Efimova}, {Ehgamberdiev}, {G{\'o}mez},
  {Gonz{\'a}lez-Morales}, {Hiriart}, {Ibryamov}, {Jadhav}, {Jorstad}, {Joshi},
  {Kadenius}, {Klimanov}, {Kohli}, {Konstantinova}, {Kopatskaya}, {Koptelova},
  {Kimeridze}, {Kurtanidze}, {Larionova}, {Larionova}, {Ligustri}, {Lindfors},
  {Marscher}, {McBreen}, {McHardy}, {Metodieva}, {Molina}, {Morozova},
  {Nazarov}, {Nikolashvili}, {Nilsson}, {Okhmat}, {Ovcharov}, {Panwar},
  {Pasanen}, {Peneva}, {Phipps}, {Pulatova}, {Reinthal}, {Ros}, {Sadun},
  {Schwartz}, {Semkov}, {Sergeev}, {Sigua}, {Sillanp{\"a}{\"a}}, {Smith},
  {Stoyanov}, {Strigachev}, {Takalo}, {Taylor}, {Thum}, {Troitsky}, {Valcheva},
  {Wehrle}, \& {Wiesemeyer}}]{Raiteri13}
{Raiteri}, C.~M., {Villata}, M., {D'Ammando}, F., {et~al.} 2013, \mnras, 436,
  1530

\bibitem[{{Rani} {et~al.}(2014){Rani}, {Krichbaum}, {Marscher}, {Jorstad},
  {Hodgson}, {Fuhrmann}, \& {Zensus}}]{Rani14}
{Rani}, B., {Krichbaum}, T.~P., {Marscher}, A.~P., {et~al.} 2014, \aap, 571, L2

\bibitem[{{Rani} {et~al.}(2013){Rani}, {Krichbaum}, {Fuhrmann}, {B{\"o}ttcher},
  {Lott}, {Aller}, {Aller}, {Angelakis}, {Bach}, {Bastieri}, {Falcone},
  {Fukazawa}, {Gabanyi}, {Gupta}, {Gurwell}, {Itoh}, {Kawabata}, {Krips},
  {L{\"a}hteenm{\"a}ki}, {Liu}, {Marchili}, {Max-Moerbeck}, {Nestoras},
  {Nieppola}, {Quintana-Lacaci}, {Readhead}, {Richards}, {Sasada}, {Sievers},
  {Sokolovsky}, {Stroh}, {Tammi}, {Tornikoski}, {Uemura}, {Ungerechts},
  {Urano}, \& {Zensus}}]{Rani13}
{Rani}, B., {Krichbaum}, T.~P., {Fuhrmann}, L., {et~al.} 2013, \aap, 552, A11

\bibitem[{{Ravasio} {et~al.}(2003){Ravasio}, {Tagliaferri}, {Ghisellini},
  {Tavecchio}, {B{\"o}ttcher}, \& {Sikora}}]{Ravasio03}
{Ravasio}, M., {Tagliaferri}, G., {Ghisellini}, G., {et~al.} 2003, \aap, 408,
  479

\bibitem[{{Richards} {et~al.}(2011){Richards}, {Max-Moerbeck}, {Pavlidou},
  {King}, {Pearson}, {Readhead}, {Reeves}, {Shepherd}, {Stevenson},
  {Weintraub}, {Fuhrmann}, {Angelakis}, {Zensus}, {Healey}, {Romani}, {Shaw},
  {Grainge}, {Birkinshaw}, {Lancaster}, {Worrall}, {Taylor}, {Cotter}, \&
  {Bustos}}]{Richards11}
{Richards}, J.~L., {Max-Moerbeck}, W., {Pavlidou}, V., {et~al.} 2011, \apjs,
  194, 29

\bibitem[{{Sambruna} {et~al.}(1999){Sambruna}, {Ghisellini}, {Hooper},
  {Kollgaard}, {Pesce}, \& {Urry}}]{Sambruna99}
{Sambruna}, R.~M., {Ghisellini}, G., {Hooper}, E., {et~al.} 1999, \apj, 515,
  140

\bibitem[{{Smith} {et~al.}(2002){Smith}, {Tucker}, {Kent}, {Richmond},
  {Fukugita}, {Ichikawa}, {Ichikawa}, {Jorgensen}, {Uomoto}, {Gunn}, {Hamabe},
  {Watanabe}, {Tolea}, {Henden}, {Annis}, {Pier}, {McKay}, {Brinkmann}, {Chen},
  {Holtzman}, {Shimasaku}, \& {York}}]{Smith02}
{Smith}, J.~A., {Tucker}, D.~L., {Kent}, S., {et~al.} 2002, \aj, 123, 2121

\bibitem[{{Smith} {et~al.}(2009){Smith}, {Montiel}, {Rightley}, {Turner},
  {Schmidt}, \& {Jannuzi}}]{Smith09}
{Smith}, P.~S., {Montiel}, E., {Rightley}, S., {et~al.} 2009, ArXiv e-prints,
  arXiv:0912.3621

\bibitem[{{Stern} \& {Poutanen}(2008)}]{Stern08}
{Stern}, B.~E., \& {Poutanen}, J. 2008, \mnras, 383, 1695

\bibitem[{{Tagliaferri} {et~al.}(2015){Tagliaferri}, {Ghisellini}, {Perri},
  {Hayashida}, {Balokovi{\'c}}, {Covino}, {Giommi}, {Madejski}, {Puccetti},
  {Sbarrato}, {Boggs}, {Chiang}, {Christensen}, {Craig}, {Hailey}, {Harrison},
  {Stern}, \& {Zhang}}]{Tagliaferri15}
{Tagliaferri}, G., {Ghisellini}, G., {Perri}, M., {et~al.} 2015, \apj, 807, 167

\bibitem[{{Ter{\"a}sranta} {et~al.}(1998){Ter{\"a}sranta}, {Tornikoski},
  {Mujunen}, {Karlamaa}, {Valtonen}, {Henelius}, {Urpo}, {Lainela}, {Pursimo},
  {Nilsson}, {Wiren}, {Laehteenmaeki}, {Korpi}, {Rekola}, {Heinaemaeki},
  {Hanski}, {Nurmi}, {Kokkonen}, {Keinaenen}, {Joutsamo}, {Oksanen},
  {Pietilae}, {Valtaoja}, {Valtonen}, \& {Koenoenen}}]{Terasranta98}
{Ter{\"a}sranta}, H., {Tornikoski}, M., {Mujunen}, A., {et~al.} 1998, \aaps,
  132, 305

\bibitem[{van~der Walt {et~al.}(2011)van~der Walt, Colbert, \&
  Varoquaux}]{numpy11}
van~der Walt, S., Colbert, S.~C., \& Varoquaux, G. 2011, Computing in Science
  Engineering, 13, 22

\bibitem[{Waskom {et~al.}(2014)Waskom, Botvinnik, Hobson, Cole, Halchenko,
  Hoyer, Miles, Augspurger, Yarkoni, Megies, Coelho, Wehner, cynddl, Ziegler,
  diego0020, Zaytsev, Hoppe, Seabold, Cloud, Koskinen, Meyer, Qalieh, \&
  Allan}]{seaborn14}
Waskom, M., Botvinnik, O., Hobson, P., {et~al.} 2014, seaborn: v0.5.0 (November
  2014), , , doi:10.5281/zenodo.12710

\bibitem[{{Wehrle} {et~al.}(2016){Wehrle}, {Grupe}, {Jorstad}, {Marscher},
  {Gurwell}, {Balokovi{\'c}}, {Hovatta}, {Madejski}, {Harrison}, \&
  {Stern}}]{Wehrle16}
{Wehrle}, A.~E., {Grupe}, D., {Jorstad}, S.~G., {et~al.} 2016, \apj, 816, 53

\bibitem[{{Wu} {et~al.}(2009){Wu}, {Gu}, \& {Jiang}}]{Wu09}
{Wu}, Z.-Z., {Gu}, M.-F., \& {Jiang}, D.-R. 2009, Research in Astronomy and
  Astrophysics, 9, 168

\bibitem[{{Zhang} {et~al.}(2014){Zhang}, {Chen}, \& {B{\"o}ttcher}}]{Zhang14}
{Zhang}, H., {Chen}, X., \& {B{\"o}ttcher}, M. 2014, \apj, 789, 66

\end{thebibliography}


\end{document}